\documentclass[lettersize,journal]{IEEEtran}
\usepackage{amsmath,amsfonts}
\usepackage{array}
\usepackage[caption=false,font=normalsize,labelfont=sf,textfont=sf]{subfig}
\usepackage{textcomp}
\usepackage{stfloats}
\usepackage{url}
\usepackage{verbatim}
\usepackage{graphicx}
\usepackage{cite}
\usepackage{cite}
\usepackage{amsmath,amssymb,amsfonts}
\usepackage[ruled,vlined]{algorithm2e}
\usepackage{graphicx,color}
\usepackage{textcomp}
\usepackage{xcolor}
\usepackage{hyperref}

\usepackage[normalem]{ulem}
\useunder{\uline}{\ul}{}
\usepackage{longtable}
\usepackage{multirow}
\usepackage{enumitem}
\usepackage{cleveref}
\crefformat{section}{\S#2#1#3} 
\crefformat{subsection}{\S#2#1#3}
\crefformat{subsubsection}{\S#2#1#3}
\usepackage{arydshln}
\setlist[itemize]{leftmargin=*}
\usepackage{wrapfig}
\usepackage{xcolor}

\definecolor{red}{rgb}{1,0,0}
\definecolor{green}{rgb}{0.1,0.55,0}
\definecolor{blue}{rgb}{0,0,1}

\newcommand{\remove}[1]{{}}

\newcommand{\add}[1]{{\color{black} #1}}
\newcommand{\update}[1]{}
\newcommand{\rev}[1]{{\color{black} #1}}
\newcommand{\revagain}[1]{{\color{black} #1}}

\newcommand{\sysname}{CANShield\xspace}

\newcommand{\defeq}{\overset{\text{\tiny def}}{=}}

\begin{document}

\title{\sysname: Deep Learning-Based Intrusion Detection Framework for Controller Area Networks at the Signal-Level}
\author{Md~Hasan~Shahriar,~\IEEEmembership{Student~Member,~IEEE,}
\thanks{M. Shahriar is with Virginia Tech, Arlington,
VA, 22203 USA e-mail: hshahriar@vt.edu.}
Yang~Xiao,~\IEEEmembership{Member,~IEEE,}
Pablo Moriano,~\IEEEmembership{Senior Member,~IEEE,}\thanks{\scriptsize{This manuscript has been co-authored by UT-Battelle, LLC, under contract DE- AC05-00OR22725 with the US Department of Energy (DOE). The US government retains and the publisher, by accepting the article for publication, acknowledges that the US government retains a nonexclusive, paid-up, irrevocable, worldwide license to publish or reproduce the published form of this manuscript, or allow others to do so, for US government purposes. DOE will provide public access to these results of federally sponsored research in accordance with the DOE Public Access Plan ({http://energy.gov/downloads/doe-public-access-plan}).}}
Wenjing~Lou,~\IEEEmembership{Fellow,~IEEE,}
and~Y. Thomas~Hou,~\IEEEmembership{Fellow,~IEEE}
\thanks{Copyright (c) 2023 IEEE. Personal use of this material is permitted. However, permission to use this material for any other purposes must be obtained from the IEEE by sending a request to pubs-permissions@ieee.org.}
\thanks{--------------- \\A version of this paper is accepted by IEEE Internet of Things Journal.}

}

\markboth{IEEE INTERNET OF THINGS JOURNAL,~Vol.~xx, No.~x, August~2023}%
{Shahriar \MakeLowercase{\textit{et al.}}: \sysname: Deep Learning-Based Intrusion Detection Framework for Controller Area Networks at the Signal-Level}


\maketitle
\begin{abstract}
Modern vehicles rely on a fleet of electronic control units (ECUs) connected through controller area network (CAN) buses for critical vehicular control. With the expansion of advanced connectivity features in automobiles and the elevated risks of internal system exposure, the CAN bus is increasingly prone to intrusions and injection attacks. As ordinary injection attacks disrupt the typical timing properties of the CAN data stream, rule-based intrusion detection systems (IDS) can easily detect them. However, advanced attackers can inject false data to the {signal/semantic level}, while looking innocuous by the pattern/frequency of the CAN messages. The rule-based IDS, as well as the anomaly-based IDS, are built merely on the sequence of CAN messages IDs or just the binary payload data and are less effective in detecting such attacks. Therefore, to detect such intelligent attacks, we propose CANShield, a deep learning-based signal-level intrusion detection framework for the CAN bus. CANShield consists of three modules: a data preprocessing module that handles the high-dimensional CAN data stream at the signal level and parses them into time series suitable for a deep learning model; a data analyzer module consisting of multiple deep autoencoder (AE) networks, each analyzing the time-series data from a different temporal scale and granularity, and finally an attack detection module that uses an ensemble method to make the final decision. Evaluation results on two high-fidelity signal-based CAN attack datasets show the high accuracy and responsiveness of CANShield in detecting advanced intrusion attacks. 
\end{abstract}
\section{INTRODUCTION}
\label{introduction}
\IEEEPARstart{M}{odern} vehicles are becoming increasingly computerized to ensure driver's safety and convenience. The fusion of multimodal data from different types of sensors enables vehicles to recognize the driving context and make crucial decisions. The majority of the vehicles' critical functionalities, including acceleration, braking, steering, engine control, etc., involve dedicated microcontroller modules, known as electronic control units (ECUs), which are connected by one or more automotive communication buses running standardized protocols. 
Controller area network (CAN), also known as the CAN bus protocol, is the de facto automobile communication standard for safety-critical ECUs~\cite{el2020cybersecurity}.
More recently, CAN bus enables vehicles to implement advanced driver assistance systems (ADAS), one of the fastest-growing applications in the automotive sector, providing enhanced passenger experience and safety. Moreover, advancements in wireless communication technology (e.g., 5G and V2X) have enabled the interface to connect with the internal ECUs from the outside network to conduct diagnostics or update firmware over-the-air (FOTA) remotely, rather than visiting a service facility~\cite{andrade2017managing}. Infotainment features such as Bluetooth, Wi-Fi, and other smart interfaces are also becoming prevalent in automobiles to add more convenience to the passengers~\cite{el2020cybersecurity}. \rev{Besides, the integration of Internet of Things (IoT) technology in the automotive industry, also known as Automotive IoT presents huge opportunities~\cite{kumar2021bdtwin}, such as optimizing the vehicles' performance, improving transportation management, and enhancing vehicle safety through predictive maintenance, AI-powered driving assistance, connectivity, etc.}

The increased connectivity of modern vehicles \rev{as well as Automotive IoT technologies} nonetheless increases the susceptibility of vehicular systems to remote attacks and message injections. The ability to hijack an ECU and inject stealthy messages into the vehicles' internal communication systems allows attackers to  circumvent a wide array of safety-critical systems and control a wide range of vehicular functions. 
Researchers discovered several remote access points on connected vehicles and demonstrated that attackers could remotely exploit them to take control of the vehicles, including disabling the brakes, braking individual wheels, stopping the engine, and so on~\cite{koscher2010experimental, woo2014practical}. 
For instance, Miller and Valasek remotely compromised a Jeep and transmitted malicious CAN messages, which led to the vehicle malfunctioning on the highway~\cite{miller2019lessons}. Later, Chrysler recalled 1.4 million vehicles that can be remotely hacked over the Internet~\cite{chrysler}.


Despite the CAN protocol's widespread implementation and high reliability, it remains vulnerable to intruders due to the absence of basic security mechanisms as they introduce delays in message transmission or increase bus traffic~\cite{jo2021survey}. Although there are a few works on implementing message authentication code (MAC) on the CAN bus to authenticate the sender ECU and prevent different attacks, they are costly and only achieve limited cryptographic strength~~\cite{xiao2020session,  schmandt2017mini}. Moreover, it is difficult to insert the MAC along with the CAN message because of the limited payload length. 
As a result, only the plaintext message is broadcast over the CAN bus. Hence, CAN protocol does not include a way to verify where the message comes from or its integrity~\cite{jo2021survey}. 
Due to this security deficiency, vehicles using the  
CAN protocol remains insecure, and attackers could, for instance, instigate sudden braking or acceleration, rendering the lives of passengers and pedestrians at risk~\cite{miller2019lessons}.

In response, an intrusion detection system (IDS) is usually regarded as the second (and most practical) line of defense, given that an attacker can hack into the vehicle's internal communication. 
In general, there are two types of vehicular IDSs---signature-~\cite{jin2021signature, olufowobi2019saiducant} and anomaly-based~\cite{halder2020coids, wu2019survey}.
A signature-based IDS typically formulates detection rules based on the system's normal behavior and known attacks. Any violations of these rules are regarded as anomalies. In CAN bus, these rules can be based on the frequency of the messages, sequence of message IDs, inter-frame time differences, signal values, etc. 
High-dimensional CAN data flow, such as broadcasting different signals/IDs at different frequencies, makes it difficult for the models to extract the effective rules~\cite{hanselmann2020canet}. Moreover, due to the limitations in the rules, these IDSs tend to show a high false-negative rate in detecting advanced attacks and, thus, require frequent updates of the known-attack database as they are only effective against known attack footprints~\cite{wu2019survey}.
Moreover, a clever attacker can even keep the sequences of the malicious CAN message benign by turning off the actual ECU through a well-known bus-off attack~\cite{cho2016error, bloom2021weepingcan} and sending crafted messages simultaneously on behalf of the victim ECU. Although a few of the works on ECU fingerprinting~\cite{choi2018identifying, cho2016fingerprinting} provided potential ways to verify the source of the CAN message by analyzing the physical layer attributes of the ECU and detecting such impersonation attacks, the assumption of the uniqueness of such physical properties is proven invalid by a recent study~\cite{bhatia2021evading}. 
\rev{Moreover, an attacker can also remotely manipulate CAN messages at the data link layer, bypassing the protocol's rules and enabling stealthy link-layer attacks~\cite{de2022canflict}. Some attacks are even possible due to 
the limitations in the physical layer\cite{mohammed2022physical}, such as different sample-point settings of ECUs~\cite{yue2021cancloak}.} 
Therefore, only analyzing the sequence of the CAN messages is not sufficient for the IDS. Rather, the only effective way to detect advanced masquerade attacks, including injection attacks, is to analyze the payload of the messages and check for abnormalities within their contents.

The second category of CAN IDS analyzes anomalies in the CAN data frame. The message IDs and the binary payloads are the main sources of data utilized in such IDSs~\cite{lokman2019intrusion}. Despite the notable advancement in anomaly-based CAN IDS research in recent years, it is still significantly hampered by several factors~\cite{verma2020road}. 
Firstly, CAN message in light-duty vehicles are obfuscated by the original equipment manufacturers (OEMs) for security and privacy reasons. Different vehicle models encode their signals using different semantic rules, even under the same OEM. 
Furthermore, in passenger vehicles, a single payload usually contains several signals, even encoded in different formats, along with some unused bits~\cite{verma2020can}. Due to this semantic gap, the anomaly-based IDSs built directly on such obfuscated complex binary CAN payloads tend to suffer high false-positive rates and lack of explainability. 

\add{Besides, any machine learning (ML)-based IDS running on raw payload data will have challenges if needing to scale with the CAN FD (flexible data-rate) technology where the payload field can be 512 bits long (instead of 64 bits)~\cite{shin2016can}.}

On the other hand, the conversion of high-dimensional binary payload data to decimal signals has several benefits~\cite{verma2020road}. First, it reduces the dimensionality of the data as many bits are combined into a single physically meaningful number. Further, it reduces the inherent noise of the binary bits, which may seem patternless cryptic fluctuations in the raw data but becomes meaningful if appropriately decoded.

Therefore, to achieve a more robust and semantically concise defense against CAN intrusions, it is imperative to design IDS schemes at the signal-level, instead of only focusing on the temporal/ID patterns and binary payload. Meanwhile, there are very few concrete proposals for the signal-level CAN IDS~\cite{hanselmann2020canet, kukkala2020indra, kukkala2021latte, moriano2022detecting}. 
Most of these considered individual deep learning models per CAN ID to track the associated time-series signals, making them impractical for modern vehicles with many CAN IDs. Moreover, as these IDSs have attack-specific designs, they lack a comprehensive detection performance against diverse types of attacks. 






Thus, in this paper, 
we propose a deep learning-based intrusion detection framework, \sysname, which can handle high-dimensional vehicular CAN bus data at the signal-level and detect advanced and stealthy attacks, including fabrication, suspension, and masquerading attacks with high accuracy and responsiveness. This framework working at the signal-level also adds transparency to the detection process. 


We make the following contributions to this paper:

\begin{itemize}
    \item We propose a deep learning-based intrusion detection framework, \sysname, to detect advanced and stealthy attacks from  signal-level CAN data. It features a data processing technique (pipeline) for the high dimensional CAN signal stream by creating a temporary data queue and using the forward filling mechanism to fill the missing data. This pipeline prepares the data stream suitable for the training and testing in the ML-based IDS. 

    \item To make the detection effective on multidimensional time series data of different temporal scales,
    we convert the two-dimensional data queues to multiple images and consider the detection as a computer vision-like problem. We consider multiple convolution neural network (CNN)-based autoencoder (AE) models to learn the various temporal (short-term and long-term) and spatial (signal-wise) dependencies. Violations in either the temporal or spatial pattern can be detected during the reconstruction process. Such data preprocessing avoids the need for individual ML models per CAN ID. 
    \item We propose a three-step analysis of the structured reconstruction loss of \sysname's AE models on the selection of detection thresholds for the optimal accuracy, 
    followed by an ensemble-based detector that boosts the overall detection performance by combining the insights from all the AEs. \rev{We also utilized transfer learning to reduce the cost of training multiple AE and ensure transferability.}
    \item  We evaluate \sysname against advanced signal-level attacks using SynCAN~\cite{hanselmann2020canet} and ROAD~\cite{verma2020road} datasets and compare the results with a baseline model to show the improvements. The results show high effectiveness and responsiveness of \sysname against a wide range of fabrication, masquerade, and suspension attacks on the CAN bus. We also make the source code publicly available\footnote{\url{https://github.com/shahriar0651/canshield}}.
    
\end{itemize}

The rest of the paper is organized as follows: We introduce necessary background information in~\cref{sec:background_motivation}. An overview of the proposed CANShield framework and the attack model is presented in~\cref{sec:system}. The technical details are shown in~\cref{sec:technical_details}.
We provide an experimental setup and implementation details in~\cref{sec:experiments}. The evaluation results are analyzed in~\cref{sec:results}. The related works are discussed in~\cref{sec:related_work}. Finally, we conclude the paper in~\cref{sec:conclusion}. 

\section{PRELIMINARIES}
\label{sec:background_motivation}


\subsection{Controller Area Network}
Robert Bosch GmbH introduced controller area network (CAN) as an automotive communication bus with the latest version (2.0) released in 1991~\cite{di2012understanding}. 

\begin{figure}[t]
    \centering
    \includegraphics[width = 0.35\textwidth]{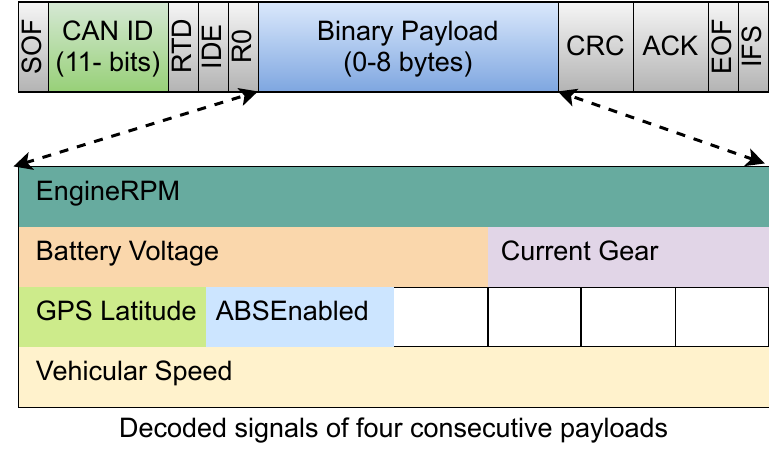}
    \caption{(Top) CAN data frame syntax. (Bottom) An example of the decoded signals that are encoded in the data field of four consecutive messages.}
    \label{fig:can_dataframe}
\end{figure}

\textbf{CAN Frame Format.~} A CAN message frame falls into four types: data frame, remote frame, error frame, and overload frame, with data frame being the default mode for data transmission. The top portion of Fig.~\ref{fig:can_dataframe} illustrates the data frame format of CAN. 
CAN data frame supports up to 8 bytes of payloads with 11 bits of arbitration ID (CAN ID), which can be extended to 29 bits.
Every ECU broadcasts its message to the CAN bus. However, only one ECU can transmit at a time and the rest stay synchronized to receive the data correctly. 
The message arbitration mechanism detects and resolves collisions of messages. A message with a higher priority contains a lower binary-encoded CAN ID. 
When any ECU detects a higher priority transmission during arbitration, it waits until the end of that message, and the channel is available to use. Due to different priorities, different CAN IDs usually appear in the CAN bus at different frequencies.

\textbf{Signal-level Representation of CAN Data.~}
The binary payload can be decoded to the signal-level using the specific car's database for CAN (DBC) file~\cite{pese2019librecan}. 
The DBC file is a proprietary format, which is quite challenging to get. However, any reverse engineering-based CAN decoder, such as the CAN-D~\cite{verma2020can}, can provide an approximate DBC file. 
Such decoding converts the binary payloads to real-valued signals and gives a time series representation. We define the time of each signal appearance as one time step. Thus, there is one CAN message at each time step, which may contain one or more associated signals along with some unused bits. The lower part of Fig.~\ref{fig:can_dataframe} shows some samples of signal-level representation of a few consecutive payloads. To prepare data input to an ML-based detector, a straightforward idea is to create a structured representation of such data stream 
, where the columns indicate different signals and rows show each time step. 
As such a data structure contains many missing entries~\cite{hanselmann2020canet}, it cannot be directly fed to the ML-based IDS models. Thus, designing an appropriate data preprocessing pipeline to account for the missing signal entries is one of the critical challenges in building a signal-level CAN IDS, as we will address in \cref{sec:filling-forward}.

\subsection{Autoencoder}
\label{sec:autoencoder}

Autoencoder (AE) is an artificial neural network that can learn efficient codings of input data through unsupervised learning~\cite{berman2019survey}. It consists of two parts: an encoder that maps an input to a lower-dimensional code and a decoder that reconstructs the closest form of the input from that code. In the reconstruction step, encoding parameters are refined so that the decoder can recover the data while retaining only the most relevant features. 
Hence, a bottleneck in the middle of the network can determine the estimated states of the system in a lower dimension. 
Let us define the function of encoder and decoder as $\phi$ and $ \psi$ that takes the input $\mathcal{X}$ and $\mathcal{F}$, respectively, such that:
$$\begin{aligned}
&\phi: \mathcal{X} \rightarrow \mathcal{F}, ~~~\psi: \mathcal{F} \rightarrow \mathcal{X} \\
&\phi, \psi=\underset{\phi, \psi}{\arg \min }\|\mathcal{X}-(\psi \circ \phi) \mathcal{X}\|^{2}
\end{aligned}$$

In intrusion detection applications AE plays a vital role. An AE network is first trained on the normal data so that it learns how to reconstruct with minimum loss. \add{The fundamental hypothesis of using AE is that intrusions are sufficiently anomalous with respect to the underlying distribution of the training data so that the AE will yield a high reconstruction loss ($\|\mathcal{X}-(\psi \circ \phi) \mathcal{X}\|$), pointing to a high probability of attack.} 

\subsection{Convolutional Neural Network}
\label{sec:cnn}
Convolutional Neural Network (CNN) is a class of deep neural networks mostly used to analyze image datasets~\cite{albawi2017understanding}. The network uses small kernels or filters that slide along the input data and map the complex relationship among the features. 
CNNs can be considered the regularized versions of multilayer perceptions and takes the advantage of the hierarchical data structure. Small filters help them learn the local and straightforward patterns first and then combine them into more complicated patterns. Therefore, CNN is an extremely powerful tool with a very low degree of connectivity and complexity. 
\rev{We build the AE networks using CNN due to the observation that each view is a two-dimensional data item, and CNN is widely proven to work efficiently on 2D data with minimum complexity.} 

\subsection{Transfer Learning}
\label{sec;transfer-learning}

Transfer learning refers to reusing a model trained for one task as the starting point for another. Pre-trained deep learning models are often used as starting points for new models if they are learning similar feature spaces and are working on similar datasets. Therefore, transferring knowledge saves time and cost during the training phase of deep learning~\cite{torrey2010transfer}. Transfer learning has two basic terms: domain and task. A domain $\mathcal{D}=\{\mathcal{X}, P(X)\}$ consists of: a feature space $\mathcal{X}$ and a marginal probability distribution $P(X)$, where $X=\left\{x_1, \ldots, x_n\right\} \in \mathcal{X}$. Given a specific domain, $\mathcal{D}$, a task $\mathcal{T}=\{\mathcal{Y}, f(x)\}$ consists of two components: a label space $\mathcal{Y}$ and a predictive function $f: \mathcal{X} \rightarrow \mathcal{Y}$. The function $f$ is used to predict the corresponding label or a representation $f(x)$ of an instance $x$. This task is learned from the training data consisting of pairs $\left\{x_i, y_i\right\}$, where $x_i \in X$ and $y_i \in \mathcal{Y}$. 

Given a source domain $\mathcal{D}_S$ and learning task $\mathcal{T}_S$, a target domain $\mathcal{D}_T$ and learning task $\mathcal{T}_T$, where $\mathcal{D}_S \neq \mathcal{D}_T$, or $\mathcal{T}_S \neq \mathcal{T}_T$, transfer learning aims to help improve the learning of the target predictive function $f_T(\cdot)$ in $\mathcal{D}_T$ using the knowledge in $\mathcal{D}_S$ and $\mathcal{T}_S$. Out of different ways, one of the most common approaches is to initiate the weights of $f_T(\cdot)$ using the trained parameters of $f_S(\cdot)$. 
The idea is that the basic structure and knowledge saved in the source model is a good start for the target model; hence, initializing  $f_T(\cdot)$ with the parameters of $f_S(\cdot)$ will reduce the initial cost.  As in this work, we consider AE-based models, $f(\cdot)$ will have the function of an AE.

\section{SYSTEM MODEL}
\label{sec:system}
\begin{figure*}[t]
    \centering
    \includegraphics[width = 0.95\textwidth]{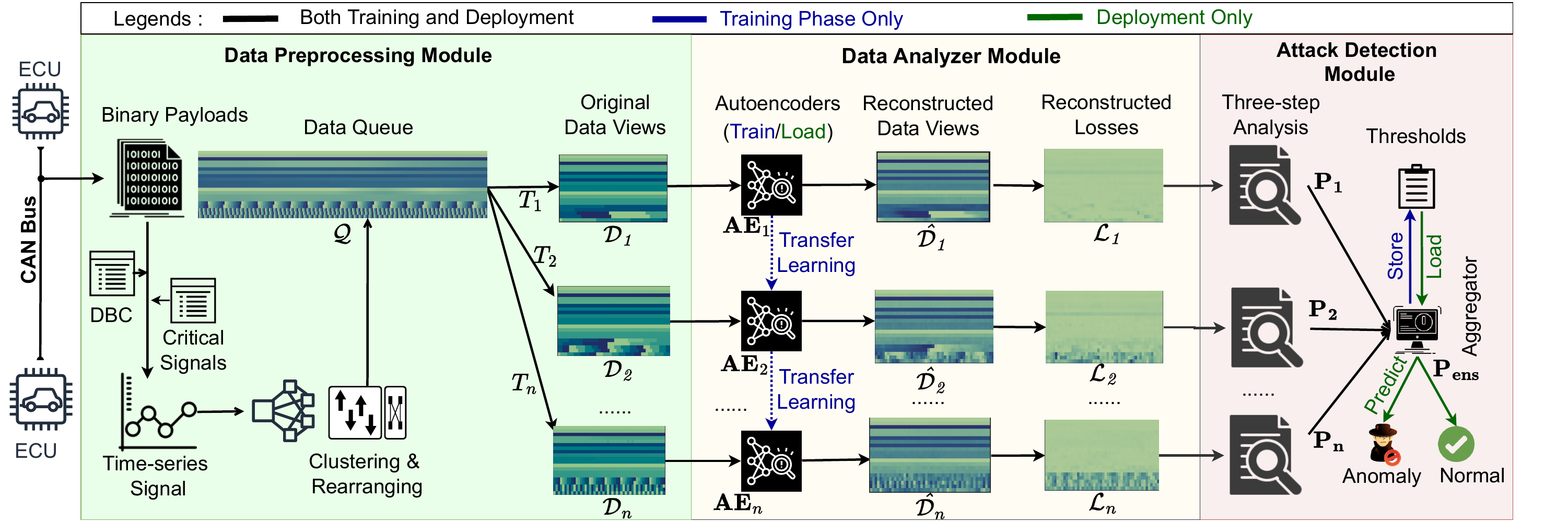}
        \caption{\sysname workflow. 
        \sysname has two phases of operation: ``{\color{blue}training}'' and ``{\color{green}deployment}''.
        \sysname contains three modules: \emph{i) data preprocessing module} that creates multiple data views of the same data queue of signal-level CAN data, \emph{ii) data analyzing module} that employs multiple CNN-based AEs for analyzing the data views and generating reconstruction losses, and \emph{iii) attack detection module} that calculates the anomaly scores and makes the final detection decision. } 
    \label{fig:ids_model}
\end{figure*}

\subsection{\sysname Overview}
The main component of \sysname is a software system that can read a vehicle's CAN messages in real-time. It is loaded either on an onboard computing device
connected to the OBD-II Port (e.g., laptop, Raspberry Pi) or instantiated in an existing ECU with a relatively powerful processor, such as the gateway ECU. For the former case, the onboard computing device includes a CAN protocol stack, allowing monitoring and recording of the raw CAN messages. This can be achieved with open-sourced implementations such as Seeed CAN-BUS Shield~\cite{seeedaduinocan} and SocketCAN \cite{SocketCAN} or commercial CAN data loggers such as CANalyzer~\cite{canalyzer}, and VehicleSpy~\cite{VehicleSpy}, etc.  
\sysname is pre-loaded with the vehicle's DBC file, either from OEM or CAN-D, allowing continuous decoding of the binary payloads, creating a data queue of multi-dimension time series signals, and tracking their changes in near real-time. 


As is shown in Fig.~\ref{fig:ids_model}, \sysname contains three modules: \emph{i) data preprocessing module} that creates multiple data views of the same data queue of signal-level CAN data, \emph{ii) data analyzing module} that employs multiple CNN-based AEs for analyzing the data views and generating reconstruction losses, and \emph{iii) attack detection module} that calculates the anomaly scores and makes the final detection decision. 
\sysname has two phases of operation: \emph{training} and \emph{deployment}.
Some of the modules play additional/slightly different roles during each of the two phases. During the training phase, the data analyzing module needs to train deep learning models. However, as the onboard devices are typically lightweight and not suitable for effective training of deep learning models, we consider two potential solutions for that. \sysname can have a secure connection to the cloud with model training capabilities or train the models on a local computer with \sysname running on that. Hence, during the training phase, the normal CAN traces are stored on the local memory first and then periodically sent to the cloud or local computer for model training. As the AEs have the same tasks (signal reconstruction) but work on slightly different domains (data views), we utilize the transfer learning technique to transfer the knowledge of one AE to the next one which is working on a higher sampling period.  Once all the models are adequately trained, \sysname loads the trained models into the onboard device and begins the deployment phase, which goes through the three modules in a feedforward fashion and outputs the detection result in near real-time. It is noted that CANShield detects attacks at the data queue level rather than at the message level.

\subsection{Attack Model}
\label{sec:attack_model}
We assume that the intruder can access the CAN bus through an exposed interface, such as V2X, infotainment, ADAS systems, OBD-II port, etc. Moreover, we also assume that the attacker is capable of turning off any ECU~\cite{cho2016error} and/or injecting arbitrarily malicious messages. \sysname is designed to protect the vehicles from the different levels of attacks in a holistic manner. In particular, according to the attacker's objective, the attacks typically fall into the following three categories:
\begin{itemize}
    \item \emph{Fabrication attacks}, wherein a compromised ECU injects malicious IDs and data to the CAN bus. However, all the legitimate ECUs are still active and also send their original data. This is the most prevalent and straightforward type of attack that is quick and easy to launch, as the attacker does not need to hijack any ECU.
    \item \emph{Suspension attacks}, wherein a legitimate ECU is turned off/incapacitated by the adversary. This attack is also called \emph{suppress attack}, where the messages from the targeted ECU disappear for a while. To achieve this, the attacker can disconnect the ECU from the in-vehicle network to prevent it from communicating. 
    \item \emph{Masquerade attacks} are the most advanced, stealthiest, and destructive attacks. This is the combination of fabrication and suspension attacks, where the attacker silences a legitimate ECU, and spoofs it in the continuing operation while injecting malicious messages. 
\end{itemize}

In evaluation, we will use a well-known CAN attack dataset, SynCAN~\cite{hanselmann2020canet} and an emergent realistic CAN dataset, ROAD~\cite{verma2020road} covering specific forms of the above attacks to test the efficacy of \sysname.

\subsection{Design Objectives}
The design objectives of the \sysname are as follows:
\begin{itemize}
    \item \textbf{Detecting advanced attacks.~} 
    The foremost objective of \sysname is to leverage established patterns and correlations of various ECU/signal states during normal driving and design a single IDS that can detect a variety of CAN message injection and manipulation attacks considered in the literature to date, particularly those advanced stealthy attacks that existing ID- or payload-based IDSs have shown ineffective in detecting. 
    
    
    \item \textbf{Near real-time detection with \revagain{low false positives}.} 
    \rev{The IDS should respond to intrusions accurately, with \revagain{low false-positive rates}, and quickly, at the same order of magnitude as the CAN message intervals, to help the vehicle avoid catastrophes.}
\end{itemize}

\section{\sysname DETAILED DESIGN}
\label{sec:technical_details}

\rev{This section elaborates on \sysname's two initializing tasks and three core modules in detail.}
\subsection{Critical Signal Selection and Clustering}
\label{sec:clustering}
As modern vehicles have hundreds of ECUs, they contain a lot of CAN IDs and numerous associated signals. Securing all of them with IDS comes with great implementation and computation costs. On the other hand, securing only a handful of important signals from the critical sub-system of the vehicle, such as the power train, engine, coolant system, etc., will reduce complexity and render feasible solutions for real-time detection.
A practical challenge arises in designing an effective detection pipeline with a select group of signals.
Accordingly, we consider \sysname to keep track of only $m$ pre-selected high-priority signals. To find the shortlisted signals, we assume that the defender has the semantic knowledge of the signals, at least on the critical signals to secure. To make the detection more effective and robust \sysname adds additional signals based on the correlation coefficient, starting from the ones with the highest correlation with the critical signals. However, adding too many signals will increase the size of the input image of the AEs, leading to an expensive and ineffective system. Therefore, $m$ is a design parameter and depends on the defender. For the rest of the paper, we will use the term ``signals" to indicate only the pre-selected $m$ signals.


The order of the signals in the created 2D input image could also impact the learning efficacy. Compared to random placement, placements that bring out stronger spatial (correlations) patterns of the signals in the resulting image will enable more effective learning. 
To facilitate the learning of the inter-sensor correlations, \sysname calculates the Pearson correlation matrix of the time-series signal dataset~\cite{benesty2009pearson}.
Interpreting the correlation coefficient as the distance between a pair of signals, \sysname utilizes hierarchical agglomerative clustering algorithm with complete linkage method~\cite{ward1963hierarchical} to find compact clusters of highly correlated signals. 
Later, we use the sequence of clustered signals to build the 2D images (queue) so that learning the signal-to-signal correlation becomes effective for the small filters of the convolutional layers. 
Therefore, if one signal starts reporting abnormal values, the CNN model will easily detect anomalies by comparing them with the nearby highly correlated signals. 
More details on the implementation are in~\cref{sec:eval-corr}. 
Notably, the two tasks, signal selection, and correlation-based clustering are done only once during the initialization of the training process (i.e., off-line with recorded data) and are not parts of the detection (deployment) pipeline. The following subsections elaborate on the three core modules of \sysname.




\subsection{Data Preprocessing Module}
\label{sec:data-preprocessing}


The data preprocessing module prepares formatted 2D inputs to the AEs of the data analyzing module. It contains the following two steps.

\subsubsection{Creating and Maintaining Data Queue}
\label{sec:filling-forward}

First of all, the data preprocessing module continuously records the CAN trace and decodes the binary payloads containing the selected $m$ signals. Then a first-in-first-out data queue $\mathcal{Q}$ is created with the historical time-series signal data for the last $q$ time steps, where $q$ is large enough for $\mathcal{Q}$ to encompass the temporal pattern of different signals. 
Thus, every new CAN message is a new entry in $\mathcal{Q}$, where the signal values only associated with that incoming CAN ID are updated.
For the rest signals, which are not updated by the new message, 
we adopt a forward-filling technique, whereas, at every time step, the missing/unreported signals are copied from the previous time step. We assume that until an ECU sends a further CAN message, its signals are still the same as the latest reported ones. Thus, as time passes, the sensor data for the last $q$ time steps are always stored in $\mathcal{Q}$.

\begin{figure}
    \centering
    \includegraphics[trim={2cm 0 0cm 0cm},clip, width = 0.45\textwidth]{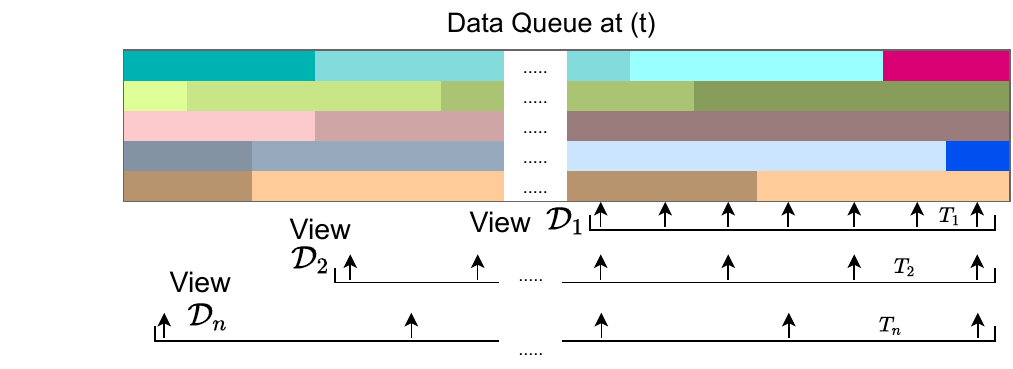}
    \caption{Generation of different views of $\mathcal{Q}$ with multiple samplings \rev{at time step $t$. For the visualization, we have transposed the original image, where the signals associated with each CAN ID are presented as a single row, and the columns indicate the time steps. The changes in the colors indicate the updates in the signal values associated with the CAN IDs. Thus, we select the first $w$ columns from $\mathcal{Q}$ at every $T_1$, $T_2$, ...., $T_n$ time steps, respectively. Here, $T_1$, $T_2$, ...., $T_n$ are the sampling periods to create the views $\mathcal{D}_1$, $\mathcal{D}_2$, ...., $\mathcal{D}_n$, respectively of the same $\mathcal{Q}$. Without the loss of generality, here we assume $T_1 < T_2 < ... < T_n$. Therefore, $\mathcal{D}_1$ has a more detailed view but contains a very limited historical trend, capturing short-term or fast-changing patterns. On the other hand, $\mathcal{D}_n$ has most of the temporal trend, capturing long-term or slow-changing patterns, but with the lowest details.}}
    \label{fig:stack_view}
\end{figure}

\subsubsection{Creating Multiple Views}
\label{sec:creating-views}
To \rev{learn the various temporal (short-term and long-term) patterns of different signals with different reporting periods and} identify abnormality, the data analyzing module needs to train and deploy the AE networks on different views \rev{(short-term and long-term)} of the data queue $\mathcal{Q}$.
As different CAN IDs have different reporting periods, only the first $w$ $(<< q)$ \rev{messages or} time steps (columns) of $\mathcal{Q}$ may not be enough to represent the recognizable temporal trend for all the signals, especially for the ones with longer reporting cycles. On the other hand, considering a high value for $w$ $(\approx q)$ makes the input image too large. As a result, the AE models become more complex. This challenge boils down to \emph{how to effectively learn the temporal pattern of all the signals, especially of the ones with long reporting periods, while still using a small time window during image generation.}

We achieve these two conflicting goals by creating $n$ different views of $\mathcal{Q}$ with $n$ different sampling periods (seeing more with a less complex model). 
Fig.~\ref{fig:stack_view} illustrates such sampling process at time step $t$ \rev{that uses sampling periods $T_1$, $T_2$, ...., $T_n$ to create the views $\mathcal{D}_1$, $\mathcal{D}_2$, ...., $\mathcal{D}_n$, respectively of the same $\mathcal{Q}$.}
The forward-filling mechanism helps to preserve the short-term or fast-changing attributes in this long-term view. Despite having different sampling periods, \rev{\sysname keeps} the number of samples ($w$) within each data view the same. As there are total $m$ signals, each data view will have a dimension of $m \times w$. This allows \sysname to use the same architecture for all the AE models \rev{working on} each data view. 
The multi-view design has benefits in the system's accuracy, 
and scalability. On the other hand, each of these views has different primary targeted signals, but collectively they cover temporal trends of variable lengths. This allows more effective and accurate detection of abnormal signals, regardless of attacking message frequency and duration. 

\subsection{Data Analyzing Module}
\label{sec:data-analyzer}
The data analyzing module utilizes multiple AE models: $\{\mathbf{AE}_i\}_{i\in[n]}$ (we define $[n]\defeq \{1,2,...,n\}$). Each of the models is associated with each of the views of $\mathcal{Q}$ and thus learns different (and complementary) perspectives of $\mathcal{Q}$. 
We build the AE networks using CNN due to the observation that each view is a two-dimensional data item, and CNN is widely proven to work efficiently on 2D data with minimum complexity. The motivation for using AE is that, as there are neither explicitly defined states of the vehicle, nor any analytical model for that, we use a data-driven approach to find the states out of a small window of the historical signal data. Thus, the data in an AE's central (bottleneck) layer represents the vehicle's state in a lower dimension. In contrast, the decoder part tries to predict the vehicle's historical signal data by looking at the state's information. If the vehicle is running in a normal state, as mostly seen in the training data, the decoder should predict accurately. Otherwise, an abnormal state will lead to an erroneous prediction, therefore, a high reconstruction loss. Moreover, as our considered model learns the relationship among all the signals, especially the nearby highly correlated ones, if at least one signal deviates from the regular pattern, \sysname will recognize it from the reconstruction loss. 

As shown in Fig. \ref{fig:ids_model}, during the training phase, each $\mathbf{AE}_x$ takes a data view $\mathcal{D}_x \in \mathbb{R}^{m \times w}$ as an input image and learns to reconstruct almost the same $\mathcal{\hat{D}}_x \in \mathbb{R}^{m \times w}$ image, $\forall x\in[n]$. Meanwhile, as \sysname trains different AEs for different views, the training cost would be linear to the number of views. Thus a practical challenge lies in \emph{how to reduce the cost of training multiple AEs}. As the views are created from the same data queue $\mathcal{Q}$, they contain inherent similarities in their structure. 

First, the spatial dependencies (correlations) along the features are still almost the same, as all the signals in each of the views are sampled with the same sampling periods. On the other hand, the temporal patterns in different views are just the expanded/shrunk versions. Hence, instead of training all the models from scratch, we consider training the first model $\mathbf{AE}_1$ thoroughly. Then we use the transfer learning technique to initialize the parameters of the next model, $\mathbf{AE}_2$, which only needs to fine-tune the parameters instead of learning everything from scratch. Thus, we initialize any $t$-th model $\mathbf{AE}_{t}$ with the preceding trained model $\mathbf{AE}_{t-1}$. Such a technique reduces the training cost (see \cref{sec:eval-tf}), which will be most effective if, in the future, the model is trained in a peripheral device like Raspberry Pi for a new vehicle.

Once the training is done, the deployment phase is initiated, and the trained models are loaded in \sysname. At the end of the training phase and during the deployment phase, the AEs are tested on the corresponding data stream and try to reconstruct the same image. For $\mathbf{AE}_x$, the absolute difference between the original image and the reconstructed image is the reconstruction loss $\mathcal{L}_x \in \mathbb{R}^{m \times w}$ is calculated as follows:

\begin{equation}
\label{eqn:recon_loss}
    \mathcal{L}_x  = abs(\mathcal{D}_x - \mathcal{\hat{D}}_x)
\end{equation}
Each element contains the corresponding signal's reconstruction loss at a certain time step, where the row and columns indicate the signal and time steps, respectively.


\subsection{Thresholds Selection and Attack Detection Module}
\label{sec:attack-detection}
In this part, we discuss how to interpret a~2D reconstruction loss $\mathcal{L}_x$ into an anomaly score $P_x$ (i.e., attack probability) for every data view $\mathcal{D}_x$ and use the results for attack detection. 

For a normal computer vision problem, the common practice would be to consider the mean value of all the elements of the absolute reconstruction loss matrix $\mathcal{L}$ as the anomaly score $P$:

\vspace{-8pt}
\begin{equation}
\vspace{-5pt}
\label{eqn-baseline}
P \leftarrow  \frac{1}{m \times w}\sum_{i =1}^{m}\sum_{j = 1}^{w} \mathcal{L}_{i,j}
\end{equation}

Compared to a normal computer vision problem, our input image (\rev{and reconstruction loss} $\mathcal{L}_x$) has a concrete structure, which gives space for tweaking the detection thresholds for better accuracy.
Thus, instead of taking the average value, we exploit the structural knowledge of $\mathcal{L}_x$ to interpret the $P_x$.
We define three types of thresholds for attack detection at each $\mathbf{AE_x}$: 
\begin{itemize}
    \item Signal-wise reconstruction loss thresholds $R_x^{Loss}\in\mathbb{R}^{m}$
    \item Signal-wise time step violation thresholds $R_x^{Time}\in\mathbb{R}^{m}$
    \item An overall signal violation threshold $R_x^{Signal}\in\mathbb{R}$
\end{itemize}

Next, we demonstrate a \emph{three-step analysis} on $\mathcal{L}_x$ to facilitate the selection of these thresholds and attack detection, as is shown in Algorithm \ref{algo:threshold-selection}, and \ref{algo:attack-detection}, respectively. 
For convenience, we have obviated the AE index $x$ for the thresholds and $\mathcal{L}$ as this approach will be applied independently to each AE. We also use three system hyper-parameters $p$, $q$, $r$ as confidence percentiles for these thresholds, which is subject to optimal tuning in practice (see \cref{sec:Hyperparameters}).

\SetKwInput{KwInput}{Input}
\SetKwInput{KwOutput}{Output}
\SetKwInput{KwVariables}{Variables}

\begin{algorithm}[t]
\scriptsize
\KwInput{Stack of reconstruction losses  $\mathcal{L}\in\mathbb{R}^{t \times m \times w}$, system hyperparameters $p,q,r$}
\KwVariables{$\mathcal{B} \leftarrow 0^{t \times m \times w}$,~~~$\mathcal{V},\mathcal{S}\leftarrow 0^{t\times m}$, }
\KwOutput{Thresholds: $R^{Loss}, R^{Time} \in \mathbb{R}^{m}, R^{Signal}\in \mathbb{R}$}

\tcc{Step 1}
%

\begin{equation}
\label{r_loss}
\forall i\in[m]:~R^{Loss}_{i} \leftarrow p^{th}~\text{\%}~\forall j, k \in[w], [t]~~\mathcal{L}^{k}_{i,j}
\end{equation}
\vspace{-2pt}

\begin{equation}
\label{binary}
\forall i,j,k \in[m], [w], [t]:~~ \mathcal{B}^{k}_{i,j} \leftarrow 1 ~~\textbf{if}~~ \mathcal{L}^{k}_{i,j} > R^{Loss}_{i}
\end{equation}

\tcc{Step 2}%
\vspace{-2pt}
\begin{equation}
\label{violation}
\forall i, k \in[m], [t]:~~ \mathcal{V}^{k}_{i} \leftarrow \sum_{j = 1}^{w}\mathcal{B}^{k}_{i,j}
\vspace{-2pt}
\end{equation}

\begin{equation}
\label{r_time}
\forall i\in[m]:~~ R^{Time}_{i} \leftarrow q^{th}~\text{\%}~\forall k\in[t]~~\mathcal{V}^{k}_{i}
\end{equation}

\tcc{Step 3}
\vspace{-2pt}
\begin{equation}
\label{signal}
\forall i, k \in[m],[t]:~~ \mathcal{S}^{k}_{i} \leftarrow 1 ~~\textbf{if}~~ \mathcal{V}^{k}_{i} >  R^{Time}_{i}
\vspace{-2pt}
\end{equation}

\begin{equation}
\label{prob}
\forall k\in[t]: P^{k} \leftarrow  \frac{1}{m}\sum_{i = 1}^{m}\mathcal{S}^{k}_{i}
\vspace{-2pt}
\end{equation}

\begin{equation}
\label{r_signal}
R^{Signal} \leftarrow r^{th}~\text{\%}~\forall k\in[t]~~P^{k}
\end{equation}
\vspace{-2pt}

 \caption{Thresholds selection for $\mathbf{AE}_x$.}
 \label{algo:threshold-selection}
\end{algorithm}
\begin{algorithm}[t]
\scriptsize
\KwInput{Reconstruction loss  $\mathcal{L}\in\mathbb{R}^{m \times w}$,\\Thresholds: $R^{Loss}, R^{Time} \in \mathbb{R}^{m}, R^{Signal}\in \mathbb{R}$}
\KwVariables{$\mathcal{B} \leftarrow 0^{m \times w}$,~~~$\mathcal{V},\mathcal{S}\leftarrow 0^{m}$}
\KwOutput{Attack decision: $attack\in \mathbb{R}$}

\tcc{Step 1}
\vspace{-2pt}
\begin{equation}
\label{binary-test}
\forall i,j \in[m], [w]:~~ \mathcal{B}_{i,j} \leftarrow 1 ~~\textbf{if}~~ \mathcal{L}_{i,j} > R^{Loss}_{i}
\end{equation}

\tcc{Step 2}
\vspace{-2pt}
\begin{equation}
\label{violation-test}
\forall i\in[m]:~~ \mathcal{V}_{i} \leftarrow \sum_{j = 1}^{w}\mathcal{B}_{i,j}
\vspace{-2pt}
\end{equation}

\begin{equation}
\label{signal-test}
\forall i \in[m]:~~ \mathcal{S}_{i} \leftarrow 1 ~~\textbf{if}~~ \mathcal{V}_{i} >  R^{Time}_{i}
\end{equation}
\tcc{Step 3}
\vspace{-2pt}
\begin{equation}
\label{prob-test}
P_x \leftarrow  \frac{1}{m}\sum_{i = 1}^{m}\mathcal{S}_{i}
\vspace{-2pt}
\end{equation}
\tcc{Ensemble}
\vspace{-2pt}
\begin{equation}
\label{prob-ens-test}
P_{ens} \leftarrow  \frac{1}{n}\sum_{x = 1}^{n} {P_{x}}
\vspace{-2pt}
\end{equation}
\begin{equation}
\label{attack-test}
attack \leftarrow 1 ~~\textbf{if}~~ P_{ens} > R^{Signal}_{ens}
\end{equation}
\normalsize
 \caption{Ensemble-based detection.} 
 \label{algo:attack-detection}
 \vspace{-2pt}
\end{algorithm}

\begin{figure}[t]
    \centering
    \includegraphics[width = 0.40\textwidth]{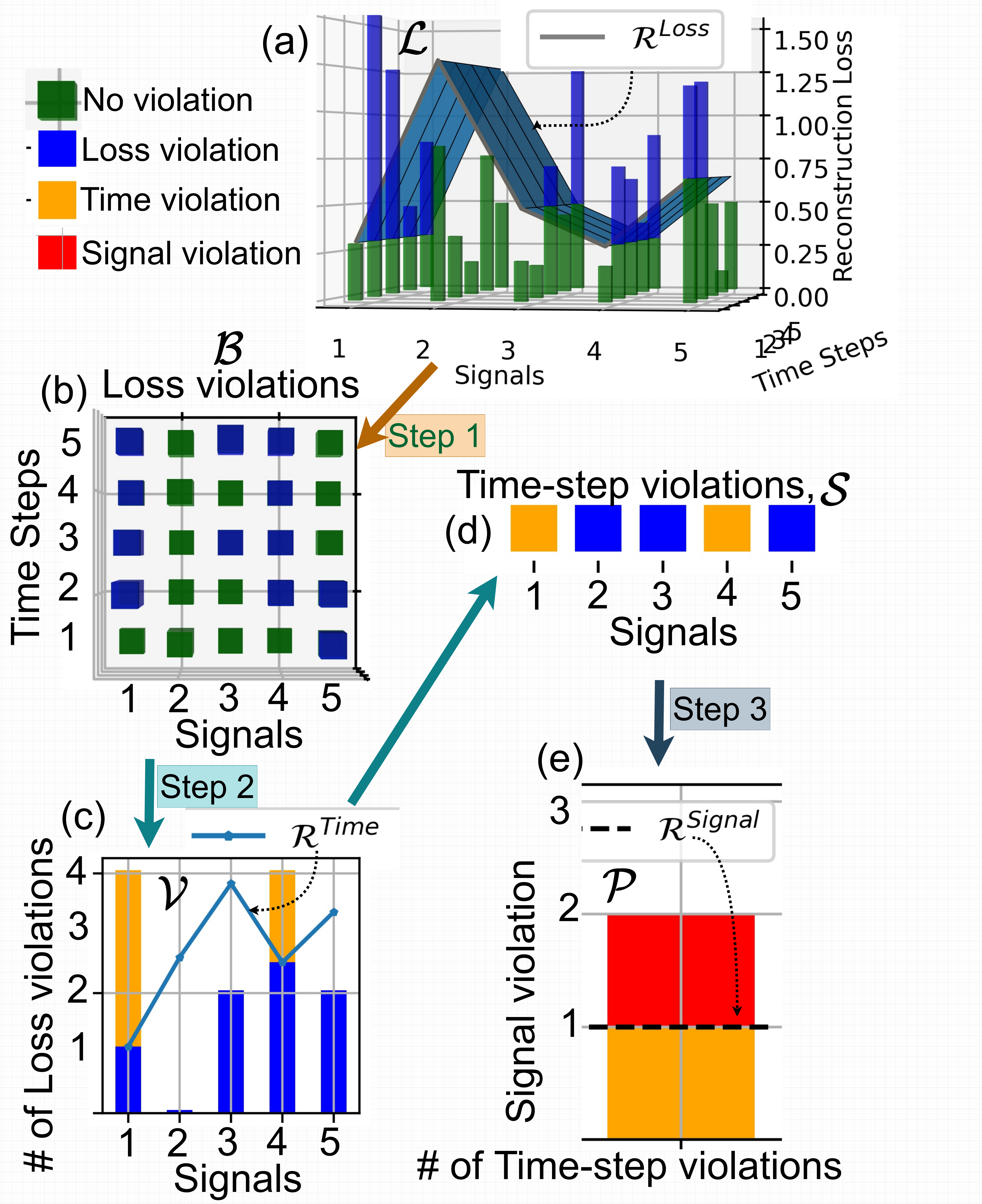}
    \caption{A simplified visual illustration of three-step attack detection (Algorithm~\ref{algo:attack-detection}) for individual AE with $5\times5$ reconstruction loss matrix.
    a) 3D visualization of 2D reconstruction loss matrix $\mathcal{L}$ showing the loss violations ($\mathcal{L} > R^{Loss}$) in blue, 
    b) Binary 2D matrix $\mathcal{B}$ showing the indices of loss violation (top view of (a)), 
    c) Signal-wise total loss violations $\mathcal{V}$ (counting only the blue bars in (b)). Orange colors show where $\mathcal{V}$ violates time-step threshold $R^{Time}$, 
    d) Binary 1D array $\mathcal{S}$ showing if any signal violates $R^{Time}$ (top view of (c)), and 
    e) Anomaly score/total signal violations $\mathcal{P}$ showing the total number of time-step violating signals (counting only the orange bars in (d)). The red color show if $\mathcal{P}$ exceeds the threshold $R^{Signal}$, indicating a potential attack; otherwise, the final prediction will be benign. For simplification, we show the total counts in the bar plots instead of using the percentage, which is used in the actual algorithm.}
    \label{fig:algo-detection}
\end{figure}

First, Algorithm \ref{algo:threshold-selection} shows how we select the thresholds from the 3D reconstruction loss matrix $\mathcal{L}$ from randomly selected $t$ training data queues. First, we find the $R^{Loss}_i$ for every signal $i\in[m]$ on the normal training data by taking the $p^{th}$ percentile values of elements in the $i^{th}$ rows of all the $\mathcal{L}$ (Eq. \eqref{r_loss}). Later, we map the 3D matrix $\mathcal{L}$ to a binary 3D matrix $\mathcal{B}$ to find the indices where the reconstruction losses are higher than the allowed threshold $R^{Loss}_i$ for every $i^{th}$ signal (Eq. (\ref{binary})). Secondly, we find the total number of such time step violations $\mathcal{V}_{i}$ for each signal by summing over all the $w$ time steps (Eq. (\ref{violation})) for all the $t$ instances. We evaluate the distribution of the signal-wise total time step violations and consider the $q^{th}$-percentile value as the time step violation threshold $R^{Time}_i$ (Eq. (\ref{r_time})).

As the third step, we check if any specific signal has more time step violations than $R^{Time}_i$ and flag that signals as compromised (Eq. (\ref{signal})) in each instance. Now we have the list of the violating signals $\mathcal{S}$ in each data view, and we consider the average value of $\mathcal{S}$ as the anomaly score $P$ for the $\mathbf{AE}$ (Eq. (\ref{prob})). Considering the false-positive requirement of the system, we set $r^{th}$ percentile value of all $P$s of the considered samples, as the total signal violation threshold $R^{Signal}$ (Eq. (\ref{r_signal})). After running all the steps, \sysname stores $R_x^{Loss}$, $R_x^{Time}$, and $R_x^{Signal}$ for each of the $\mathbf{AE_x}$, and consider the average of all $R_x^{Signal}$s as the threshold $R_{ens}^{Signal}$ for the ensemble model.

During the deployment phase, these thresholds are pre-loaded from the memory and Algorithm~\ref{algo:attack-detection} is used to detect any violation. Although the tasks in (Eqns (\ref{binary-test} - \ref{prob-test})) are similar as Algorithm~\ref{algo:threshold-selection}, \sysname runs them on individual test reconstruction loss $\mathcal{L}$ and check for potential threats using the ensemble model. Here, an anomaly score is assigned on each of the reconstruction losses on the data views, i.e., $P_1$, $P_2$, $\cdots$, $P_n$.
\sysname then uses the ensemble anomaly score $P_{ens}$ (Eq. (\ref{prob-ens-test})) as the final score. In the case of $P_{ens} > R^{Signal}_{ens}$, the IDS tags $\mathcal{Q}$ as anomalous and raises the alarm in the system (Eq. (\ref{attack-test}). Compared to the mean absolute value method (Eq. \ref{eqn-baseline}), this three-step method gives \sysname 
finer decomposition of $\mathcal{L}$ and improves the detection efficacy against stealthy attacks. Fig~\ref{fig:algo-detection} shows a simplified visualization of Algorithm~\ref{algo:attack-detection} with a $5\times5$ reconstruction loss matrix.

\section{IMPLEMENTATION AND EVALUATION}
\label{sec:experiments}

\subsection{Datasets and Attacks}
\label{sec:dataset}

We implement \sysname on both the SynCAN dataset and ROAD dataset. SynCAN dataset~\cite{hanselmann2020canet} (Synthetic CAN Bus Data) is a widely used CAN attack dataset released by ETAS (a subsidiary of Robert Bosch Gmbh) covering stealthy signal-level CAN attacks. ROAD dataset~\cite{verma2020road} was released by Oak Ridge National Laboratory and is the most realistic CAN attack dataset to date\footnote{
\rev{{To the best of our knowledge, the SynCAN dataset (available at \url{https://github.com/etas/SynCAN}) was the only publicly available signal-level CAN dataset with advanced attacks at the time of writing this paper. ROAD dataset (available at \url{https://0xsam.com/road/}) was obfuscated and did not have signal-level interpretation in its initial release in early 2021. We obtained the raw ROAD dataset by directly contacting ORNL. Partially motivated by our work, ORNL has recently released a signal-level ROAD dataset.}}}
Next, we introduce the details of each dataset and the attacks covered.

\subsubsection{SynCAN} The SynCAN dataset is built on actual CAN traces, emulating the characteristics of the real CAN traffic, with hundreds of advanced attack scenarios. It contains a total of 20 signals, including physical values, counters, and flags. 
There are 24 hours of logged data, of which 16.5 hours are for training and 7.5 hours are for testing with five types of advanced attacks, which resembles the three stealthy forms of attack models mentioned in \S\ref{sec:system}-\ref{sec:attack_model}.


 


\begin{table}[t]
\scriptsize
\caption{Description of attacks in SynCAN dataset.}
\label{tab:syncan_attacks}
\resizebox{\columnwidth}{!}{%
\begin{tabular}{|l|l|l|}
\hline
\textbf{Attack Name} & \textbf{Attack Type}  & \textbf{Description}                              \\ \hline
Flooding             & Fabrication           & Frequently injects high-priority messages.      \\ \hline
Suppress             & Suspension            & Prevent an ECU from transmission.             \\ \hline
Plateau & \multicolumn{1}{c|}{\multirow{3}{*}{Masquerade}} & Broadcasts a constant value. \\ \cline{1-1} \cline{3-3} 
Continuous           & \multicolumn{1}{c|}{} & Broadcasts continuously changing values.       \\ \cline{1-1} \cline{3-3} 
Playback             & \multicolumn{1}{c|}{} & Broadcasts a series of recorded values. \\ \hline
\end{tabular}%
}
\end{table}

The attacks in SynCAN datasets are summarized in Table~\ref{tab:syncan_attacks}.
 In a \textit{flooding attack}, the attacker frequently broadcasts high-priority messages to delay the legitimate ECUs' transmission (similar as DoS attack). In a \emph{suppress attack}, the attacker turns off the corresponding ECU of the targeted signal(s) or prevents it from sending further messages.
 Based on the time-series nature of the injected data, there are three types of masquerade attacks. In a \textit{plateau attack}, the attacker broadcasts the same constant value of any signal over a long period of time. 
 The impact of such an attack depends on the extent of the leap and the duration of the attack. In a \textit{continuous attack}, the signals are overwritten with continuously changing values that shift from the actual ones. 
 Such small changes can initially look realistic and bypass IDS. Lastly, in a \textit{playback attack}, the attacker replays a series of previously recorded data for the targeted signal to make it more realistic. 


\begin{table}[!t]
\scriptsize
\caption{Description of masquerade attacks in ROAD dataset.}
\label{tab:road_attacks}
\begin{tabular}{|l|l|}
\hline
\textbf{Attack Name} & \textbf{Description and impacts} \\ \hline
Correlated signals   & Inject different values for wheel speeds that stop the car. \\ \hline
Max speedometer      & Inject maximum value to display on the speedometer. \\ \hline
Max coolant  temp    & Inject maximum value; turn the coolant warning light on.   \\ \hline
Reverse light on/off & Toggle the reverse light that does not reflect the gear. \\ \hline
\end{tabular}
\end{table}

\subsubsection{ROAD} The ROAD dataset provides the highest-fidelity CAN traces with physically verified most realistic CAN attacks. It contains a significant amount of training data covering the different contexts of driving. We obtained the raw ROAD dataset and extracted signals from the CAN messages using CAN-D. 
There are 3.5 hours of logged data, of which 3 hours are for training and 30 minutes are for testing with five types of advanced masquerade attacks targeting the engine coolant temperature, engine RPM, brake light, and wheel speed sensors. The injected message manipulates only the specific portion of the data fields containing the targeted signals. 

Whereas the attacks in the SynCAN dataset are created by post-processing (replacing original ones) on the normal driving data, the attack traces in the ROAD traces were collected from a real vehicle under the real injection attacks. Such attack traces provide not only the injected messages but also the response from the vehicle under such attacks, which makes the ROAD dataset the most realistic one. The attacks in the ROAD dataset are summarized in Table~\ref{tab:road_attacks}. In light of the model's complexity, one single IDS is not a feasible option to track all the hundreds of decoded signals within the ROAD dataset. 
Thus, in the implementation of \sysname on the ROAD dataset, we consider seven primary signals, which were compromised during the attacks, to be of primary importance and add two highly correlated signals for each to make the IDS more robust, as detailed in \cref{sec:clustering}.

\subsection{Evaluation Setup}
\label{sec:eval_Setup}
\subsubsection{\sysname Software Implementation}
We use Python 3.7.3 with Keras 2.2.4~\cite{keras} 
for training and evaluation of CANShield. 
The pipeline for the AE model contains the combinations of the convolutional layer, activation layer (LeakyRelu), max pooling, and up-sampling layers~\cite{albawi2017understanding}. Using min-max scaling, we keep the values of each signal between 0 and 1. 
We used a five-layer network, where the convolutional layers have $3\times3$ filters, and the numbers filters in each layer are 32, 16, 16, 32, and 1. We utilized leakyRelu as the activation function with a parameter of 0.2, except for the output layer, which has a sigmoid activation function. The decoder part contains up-sampling layers with $2\times2$ filters. We use the Adam optimizer with a learning rate of 0.0002 to train the models and mean square error as the loss function. Using a batch size of 128, we train each model for 100 epochs. 
The following section explains the impact of different parameters in attack detection and illustrates the effectiveness of \sysname. 

\begin{figure}[!t]
    \centering
    \includegraphics[width = 0.40\textwidth]{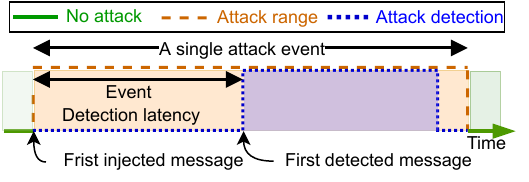}
    \caption{Attack detection and event detection latency in a single attack event.}
    \label{fig:metric}
\end{figure}
\subsubsection{Evaluation Settings}
\label{sec:views}

\rev{To evaluate \sysname, we consider $w$ as 25, 50, and 100, and five sampling periods ($T_x$) as 1, 5, 10, 20, and 50 for each of the datasets. After training the AE models, we select a random 10\% of the samples from the training data and determine the reconstruction losses using Eqn (\ref{eqn:recon_loss}) and time step violations for each AE. We also study the comparative analysis of the effectiveness of different sampling periods against different attacks. We do an extensive grid search with all the combinations of threshold ranging from 90\% to 99.99\% as $p$, $q$, and $r$ to find $R^{Loss}$, and $R^{Time}$, and $R^{Signal}$, respectively, as mentioned in Eqns (\ref{r_loss}, \ref{r_time}, and (\ref{r_signal}), to evaluate CANShield and maximize detection performance. 
Moreover, we evaluate different detection scenarios by setting 0.1\%, 0.5\%, and 1\% as the maximum threshold for the false positive rate (FPR) in the system. 

With these settings, we evaluate \sysname's performance on the following three aspects: 

\noindent\textbf{Attack detection.~} Any injection or modification of any CAN message, as is described in the attack model in \S\ref{sec:attack_model}, is considered an attack. Attack detection is defined as the detection of any malicious data view. If any view of the data queue contains one or more malicious injections, we consider the label of the queue view as malicious. 

\noindent\textbf{Event detection latency.~} Depending on the type of attack, there could be a delay between the first injected message and the first correct detection during any attack event. Such a delay is defined as the event detection latency. 
Fig.~\ref{fig:metric} shows the event detection latency for a single attack event.

\noindent \textbf{Hardware processing latency.~}
We evaluate \sysname's performance by implementing it on a standard computer as well as a lightweight edge device and benchmark the inference time, showing the \revagain{near real-time} performance in hardware. }

\subsubsection{Evaluation Metrics}

For any binary classifier, there are four possible outcomes. True positive (TP), and true negative (TN) are the outcomes where the model correctly predicts the positive (attack), and negative (benign) classes, respectively. A false positive (FP), and false negative (FN) are the outcomes where the model incorrectly predicts the positive classes, and negative classes, respectively. Based on these outcomes, we evaluate CANShield's performance using the following metrics:

\begin{itemize}

\item {\textbf{Precision}} is defined as the ratio between the correctly predicted positive data views to a total number of predicted positive views ($\frac{TP}{TP + FP}$).  
\item  \textbf{Recall} or \textit{True Positive Rate (TPR)} is calculated as the ratio between the number of positive views correctly classified as positive to the total number of actual positive views ($\frac{TP}{TP + FN}$). 


\item {\textbf{False Positive Rate (FPR)}} is the proportion of negative views incorrectly identified as positives ($\frac{FP}{FP + TN}$).  

\item {\textbf{F1 Score}} is the harmonic mean of precision and recall ($2 \times \frac{Precision \times Recall}{Precision + Recall}$). For an imbalanced dataset, F1 score is mostly used to evaluate the model's performance. 
\item {\textbf{ROC Curve, PR Curve, and AUC Scores}} indicate the classifiers performance with varying discrimination thresholds~\cite{hanley1982meaning}. 
The ROC curve plots TPRs and FPRs, and the PR curve plot precisions and recalls for different thresholds. The area under the ROC and PR curves are represented as AUROC, and AUPRC, respectively, which indicate the robustness of the detectors. 
An ideal detector has both AUROC and AUPRC scores of 1.00. 
\end{itemize}



\subsubsection{Baseline Models}
\label{sec:baseline}

\rev{We consider CANShield with only one AE with sampling periods $T_x$ as CANShield-$T_x$ and the full-fledged multi-AE-based CANShield as CANShield-Ensemble (or CANShield-Ens).} 

This part describes the four baseline models that we consider for the performance comparison.
\begin{itemize}
    \item \textbf{CANShield-Base}: We consider CANShield-Base, a simplified version of CANShield to represent the existing approaches in CNN-AE-based IDS working on windows of multi-dimensional time-series data~\cite{chen2018autoencoder}. We consider CANShield-Base to have only one AE working with a sampling period of $1$ using the conventional one-step mean absolute value of reconstruction loss (as Eqn.~\ref{eqn-baseline}) to calculate the anomaly score. \rev{Hence, the performance comparison between CANShield-Ens and CANShield-Base justifies the significance of multiple AEs and a three-step analysis of reconstruction losses.} 
    \item \textbf{CANet}: CANet~\cite{hanselmann2020canet} is the IDS specifically designed for high-dimensional CAN data structure, employing one LSTM model for each CAN IDS and merging their output to create a fully connected AE network. The authors evaluated CANet on the SynCAN dataset and made the dataset public~\cite{syncan}. As we are also utilizing the SynCAN dataset, CANet becomes the most relevant baseline for CANShield-Ens. 
    \item \textbf{Reconstructive}: The fundamental approach of the reconstructive baseline is similar to CANShield-Base. Whereas CANShield-Base feeds all the signals in one single AE model, reconstructive baseline uses different AE models for different signals~\cite{weber2018online}. Therefore, although it can learn the temporal dynamics, there is no way to learn the signal-wise correlations.
    \item \textbf{Predictive}: In the predictive baseline, there are individual LSTM models for each CAN ID that predicts the signals associated with the CAN ID for the next time-step~\cite{taylor2016anomaly}. 
    Hence, whereas all the reconstruction-based methods, including CANShield and CANet, rely on the reconstruction of the input that contains the past and current values, the predictive baseline forecasts the future values from the given input and compares them with the reported ones. 
\end{itemize}

\section{EVALUATION RESULTS AND DISCUSSION}
\label{sec:results}


This section, firstly, explains why correlation-based clustering is effective for CANShield; and later shows \sysname's performance on the different aspects. 

\begin{figure*}[t]
    \centering
        \includegraphics[width=0.85\textwidth]{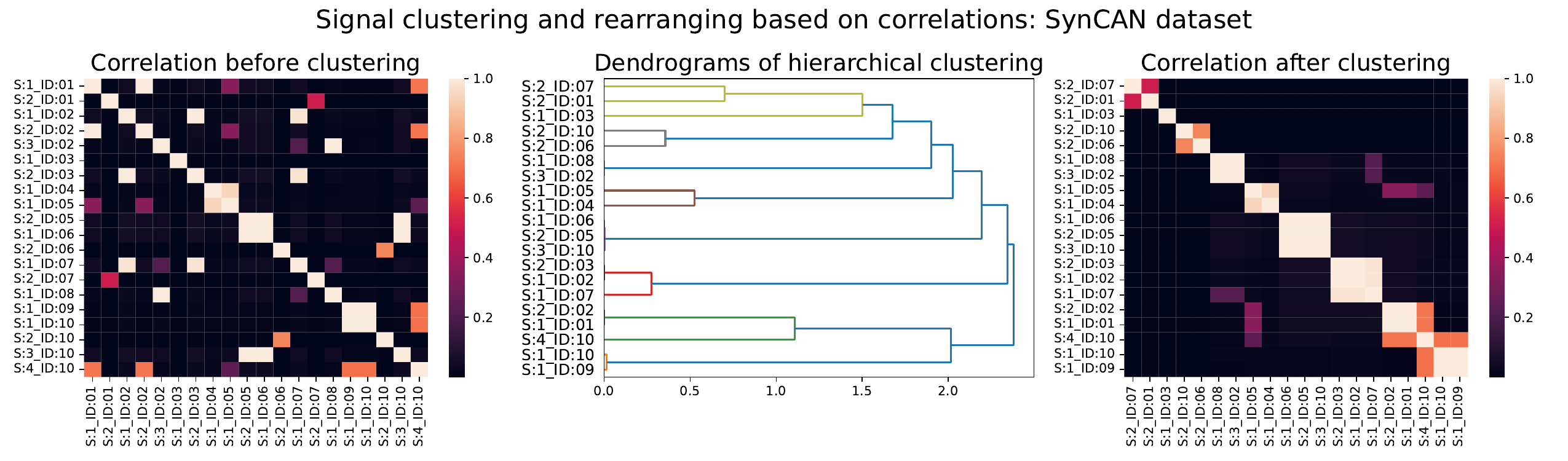}
         \caption{Hierarchical clustering of the signals in SynCAN dataset based on the correlation matrix, and rearranging them in clusters.}
    \label{fig:data_clustering}
\end{figure*}

\subsection{Correlation-based Clustering}
\label{sec:eval-corr}
As discussed in~\cref{sec:clustering}, in the initialization of the training phase, \sysname analyzes the Pearson correlations matrix of the dataset to create clusters of signals, and rearrange them so that highly correlated signals stay together in the data queue $\mathcal{Q}$. 
The left panel of Fig.~\ref{fig:data_clustering} shows the heat map of the correlation matrix of the SynCAN dataset, with the original orders of the signals as appeared in the dataset. It is clear from the figure that some of the highly correlated signal pairs, for example, S:1\_ID:02, and S:1\_ID:07, have a correlation of around unity but originally, they are placed far apart. Such placement makes it harder for the small CNN filters to learn their dependencies. 
The middle panel of Fig.~\ref{fig:data_clustering} shows the dendrograms after correlation-based clustering, which also indicates the existence of multiple clusters of highly correlated signals. 
For example, in the SynCAN dataset, S:1\_ID:10 and S:1\_ID:09 form a cluster of two signals, and S:2\_ID:03, S:1\_ID:07, and S:1\_ID:02 form another cluster. 
The right panel of the figure shows the heat map of the correlation matrix after the signal reordering. Therefore, such grouping and reordering make data queue $\mathcal{Q}$ generation more interpretable and effective. 

\begin{figure}[!t]
    \centering
    \includegraphics[width = 0.49\textwidth]{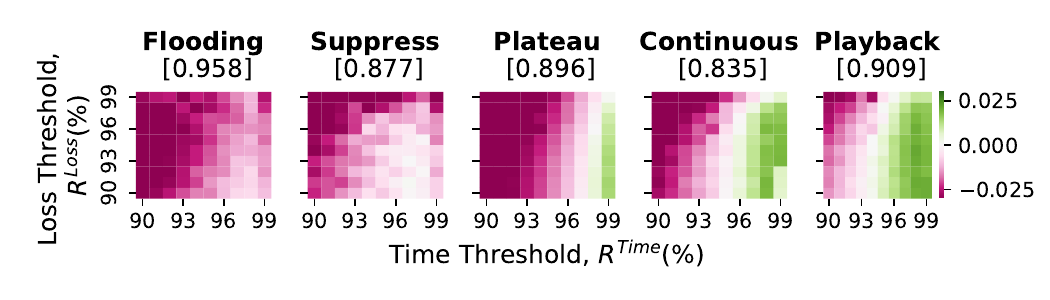}
    \caption{Effectiveness of three-step loss analysis in CANShield over the mean absolute loss in CANShield-Base. The values within the [~] show the AUROC scores of CANShield-Base, whereas the \rev{colors of the} pixels show the improvements in the AUROC scores for different ${R}^{Loss}$ and ${R}^{Time}$.}
    \label{fig:threshold_selection}
\end{figure}

\subsection{Attack Detection}
\label{sec:attack_detection}

\subsubsection{Optimizing Design Hyperparameters}
\label{sec:Hyperparameters}
We first show how we optimize \sysname's system hyperparameters to achieve the best performance on the SynCAN dataset. We assess the contribution of each feature of \sysname in attack detection in the three following steps: 



\vspace{5pt}
\noindent\textbf{Effectiveness of Three-step Analysis.~}
As the first version of CANShield, we consider CANShield-1, which uses only one AE working on a sampling period of $1$ and a data view length of $50$. Thus, the three-step analysis of reconstruction loss is the only difference between CANShield-1 and CANShield-Base. Hence, we demonstrate the efficacy of the three-step analysis of reconstruction loss (in CANShield-1) over the mean absolute loss (in CANShield-Base) by selecting different values for thresholds $R^{Loss}$, $R^{Time}$, and $R^{Signal}$, respectively. 
The captions of the sub-figures in Fig.~\ref{fig:threshold_selection} show the AUROC score of CANShield-Base for each attack type, while different pixels indicate the improvements in the AUROC scores of CANShield-1 over \sysname-Base for different combinations of $R^{Loss}$ and $R^{Time}$.



The figure shows whereas the proposed three-step analysis has limited contributions on the \textit{flooding} and \textit{suppress} attacks (first two panels), it provides a better representation of violations and improves the detection performance of the stealthy \textit{masquerade} attacks (last three panels) compared to CANShield-Base. As the violations in the \textit{fabrication} and \textit{suspension} attacks are more evident and do not involve any modification of signals, mean absolute loss itself suffices to give a decent detection performance (AUROC scores of 0.958, and 0.877, respectively). However, setting $R^{Loss}$ and $R^{Time}$ to $95$-percentile and $99$-percentile, respectively, helps better analyze the nuanced violations created by the \textit{masquerade} attacks and provides the improvements (0.02$\sim$0.03 in AUROC scores) over CANShield-Base. This evaluation shows adding a three-step analysis improves the detection rate even when one AE is used. In the following paragraphs, we will discuss how adding more AEs, and combining them into an \textit{ensemble} detector, CANShield-Ens further improves the detection performance.

\begin{figure}[!t]
    \centering
    \includegraphics[width=0.49\textwidth]{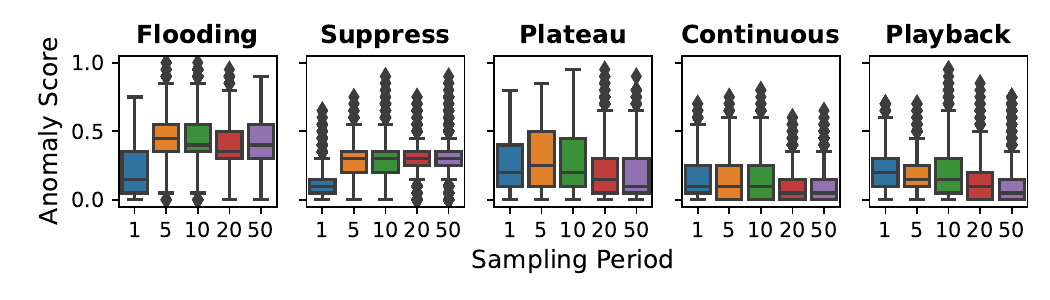}
    \caption{Anomaly scores of CANShield with different sampling periods on malicious samples. Higher anomaly scores on malicious samples make the IDS more effective.}
    \label{fig:multiple-aes}
\end{figure}

\vspace{5pt}
\noindent\textbf{Effectiveness of Different Sampling Periods.~} Here, we demonstrate the effectiveness of learning from multiple views with multiple AEs working with different sampling periods in detecting attacks. Fig~\ref{fig:multiple-aes} illustrates the performance comparison of CANShield-$T_x$, where $T_x\in \{1, 5, 10, 20, 50\}$. We analyze the effectiveness of CANShield-$T_x$ by plotting the distributions of anomaly scores of the malicious data queues. As the anomaly scores on the benign data queues are mostly zeros, we only show the anomaly scores on malicious data queues. The first two panels of the figure show that for both \textit{flooding} and \textit{suppress} attacks, the anomaly scores of the malicious data queues increase for higher sampling periods, making the detection easier as these attacks are more detectable looking at the long-term sequential pattern. 
As higher anomaly scores on malicious data queues make the classification task easier, it increases the TPR while lowering the FPR. Therefore, AE working on a higher sampling period ($\ge 5$) is the most effective against \textit{fabrication} and \textit{suspension} attacks. 

On the other hand, a sampling period of $5$ seems to be the most suitable choice against \textit{plateau} attacks, and a sampling period of $1$ is the best performing one against the \textit{continuous}, and \textit{playback} attacks. Hence, unlike \textit{fabrication} and \textit{suspension} attacks, the lower sampling periods ($\le 5$) are, in general, the most effective ones against the \textit{masquerades} attacks as short-term views of the data queue provide a detailed look at the time-series abnormalities. Therefore, only one AE working on only one type of data representation is not enough to detect diverse attacks. This finding motivates the design of \sysname-Ens, combining multiple AEs into a single decision model to further increase the robustness of the IDS.

\vspace{5pt}
\noindent\textbf{Effectiveness of Ensemble Model.~}
To design the final ensemble model, we studied different combinations of AEs working with different sizes of data views. Here, we consider the standard ensemble technique of averaging multiple anomaly scores to a single score (attack probability) (as mentioned in Eqn (\ref{prob-ens-test}) and use that to evaluate the detection performance. 

\begin{figure}[!t]
    \centering
    \includegraphics[width=0.495\textwidth]{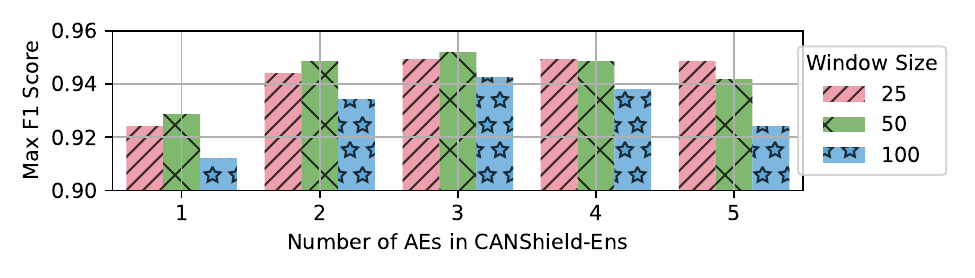}
    \caption{Optimizing \sysname-Ens's architecture. Best AUROC score for different window size $w$ ($\{25, 50, 100\}$) and AEs.}
    \label{fig:System}
\end{figure}
\begin{figure}[!t]
    \centering
  \subfloat[\scriptsize{SynCAN dataset}\label{fig:attack_detect_roc_syncan}]{%
      \includegraphics[width=0.49\linewidth]{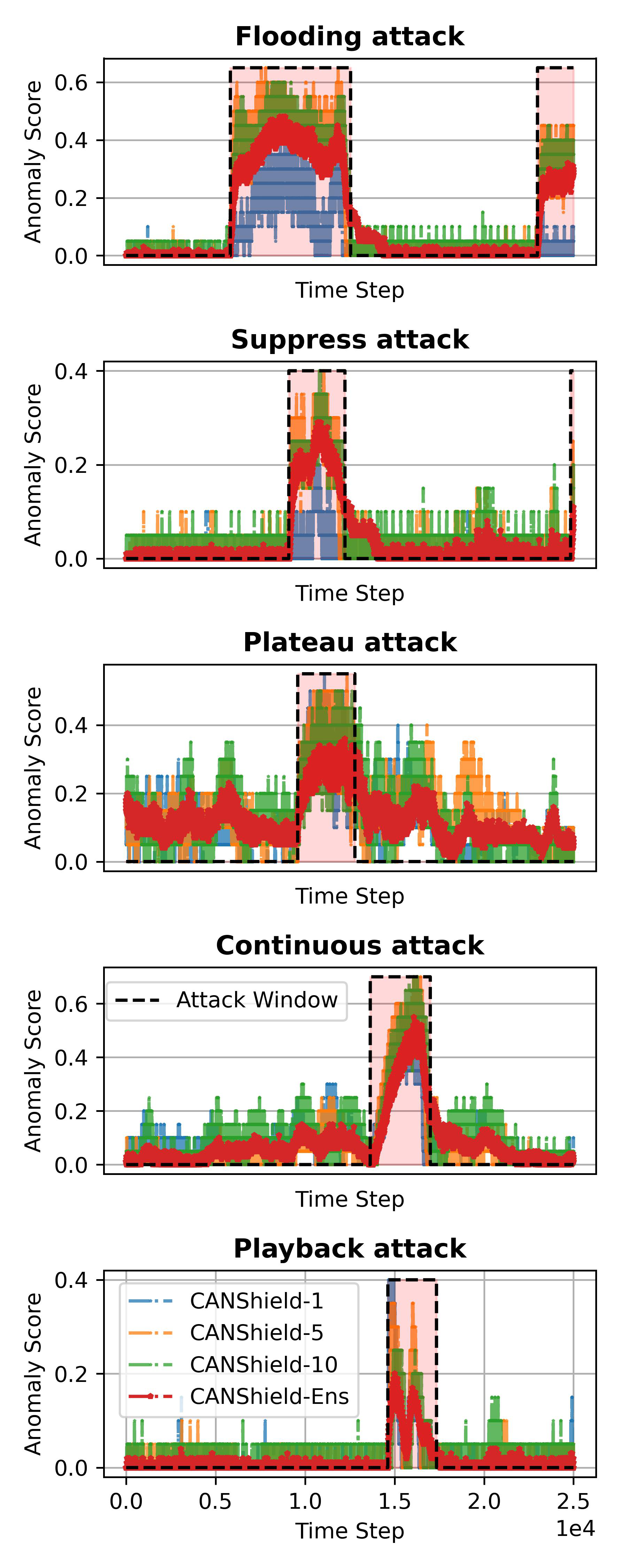}}
    \hfill
  \subfloat[\scriptsize{ROAD dataset}\label{fig:attack_detect_roc_road}]{%
        \includegraphics[width=0.49\linewidth]{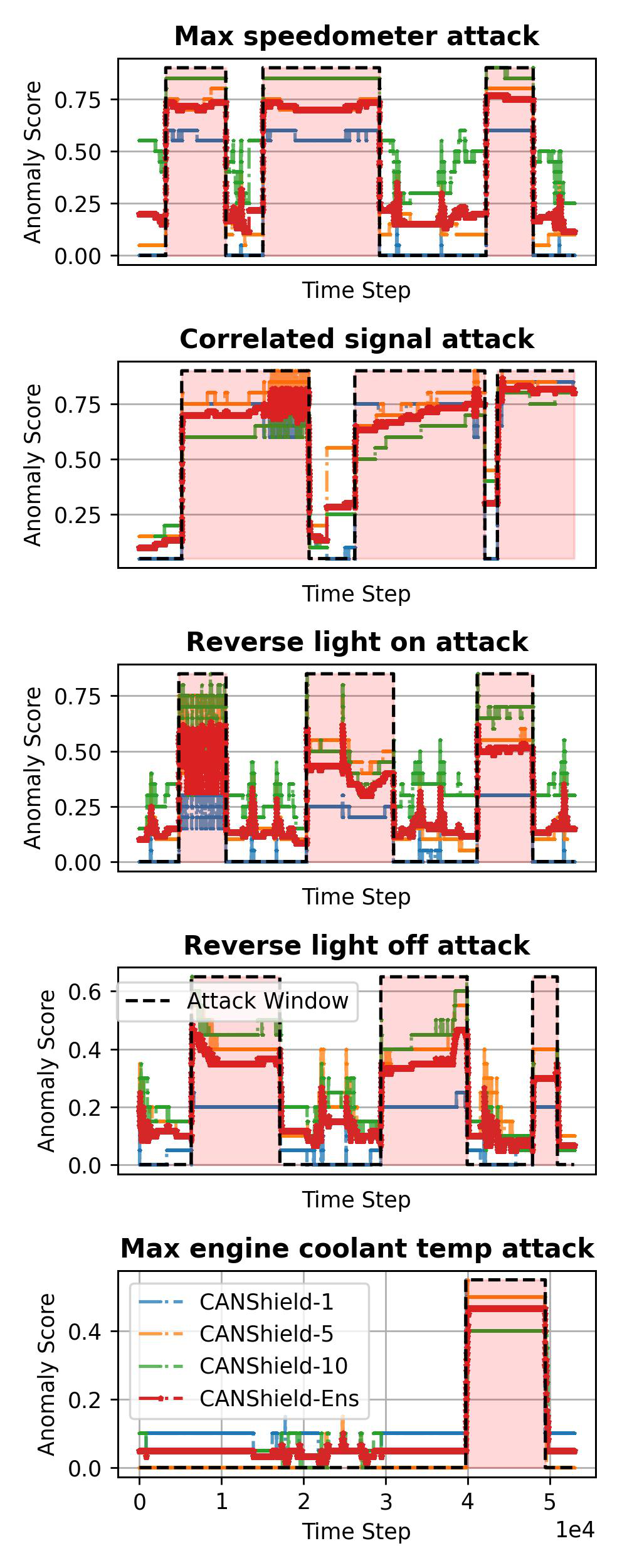}}

  \caption{Attack visualization with different models for both datasets.}
\label{fig:attack_detect_roc}
\end{figure}

To search for the final ensemble model, we studied different window sizes and different combinations of AEs, starting from one AEs up to five AEs.
As Fig.~\ref{fig:System} shows, CANShield with only one AE has limited performance (AUROC score $<0.93$) regardless of the window size $w$. When more AEs are ensembled, the performance improves. Although $w=25$ shows promising performance, it still underperforms that of $w=50$ even having more AEs. Besides, we observe that $w=100$ tends to make the model overly complicated and yield performance degradation. From the figure, it is evident that, on average, CANShield-Ens performs the best on the SynCAN dataset when $w=50$ and there are three AEs working. We further find that out of various combinations of three sampling periods, the ensemble of 1, 5, and 10 gives the best performance.  

We note that although the above results are derived from the SynCAN dataset, the ROAD dataset also shows a similar result. Therefore, for the simplicity of the analysis, we use $w$ as $50$, three AEs (with sampling periods $1$, $5$, and $10$), and $R_{Loss}$ and $R_{Time}$ as $95^{th}$-percentile and $99^{th}$-percentile, respectively, for both the SynCAN and ROAD datasets in the following evaluations.

\subsubsection{Attack visualization and AUROC Scores}
\label{sec:attackvisandroc}

In this part, we visualize the anomaly scores for all the individual and ensemble detectors along with the ROC curves for both SynCAN and ROAD datasets.

\vspace{5pt}
\noindent\textbf{SynCAN Dataset.~}
Fig.~\ref{fig:attack_detect_roc_syncan} shows the CANShield's anomaly scores, and \revagain{the left panel of} Table~\ref{tab:baseline} summarizes AUROC scores for different attacks on the SynCAN dataset. Different AEs (CANShield-$T_x$) show different performances on each of the attacks. However, the CANShield-Ens yields more stable and consistent performance, leading to higher AUROC scores in all the attacks than the individual CANShield-$T_x$.
In the case of \textit{continuous} and \textit{playback} attacks, the signals start to deviate gradually from the original values, which takes some time to create the recognizable deviation for the IDS. Hence, a lower AUROC score in the CANShield-Ens is not unexpected, especially against \textit{continuous} attacks. 
However, \sysname-Ens can detect the violations almost instantly for the rest of the attacks (AUROC scores of $0.95\sim1.00$). Whereas the individual AEs are attack-specific, the ensemble model takes the best out of every model, generalizes the process, and detects most attacks with the highest AUROC scores.

\begin{table*}[]
\centering
{\color{black}
\caption{Performance comparison with different CANShield architectures and baseline detectors on SynCAN dataset.}
\label{tab:baseline}
\resizebox{\textwidth}{!}{%
\begin{tabular}{lllllll|lllll}
\hline
 & \multicolumn{6}{c}{Area Under the ROC Curve (AUROC)} & \multicolumn{5}{c}{True Positive Rate 
 / False Positive Rate} \\\hline\hline
 & Flooding & Suppress & Plateau & Continuous & Playback & Average & Flooding & Suppress & Plateau & Continuous & Playback \\\hline
CANShield-1    & {0.940}           & 0.860           & 0.907          & 0.853          & 0.927      & 0.898        & 0.581 / 0.009 & 0.166 / 0.008 & 0.42 / 0.007 & 0.361 / 0.007 & 0.657 / 0.006\\
CANShield-5    & {0.997}  & 0.976          & 0.944          & 0.837          & 0.905      & 0.932        & 0.992 / 0.01 & 0.82 / 0.008 & 0.482 / 0.009 & 0.353 / 0.007 & 0.594 / 0.004 \\
CANShield-10   & 0.994          & 0.978          & 0.924          & 0.814          & 0.888      & 0.920       &   0.976 / 0.008 & 0.846 / 0.009 & 0.429 / 0.009 & 0.247 / 0.006 & 0.589 / 0.005 \\\hline
\sysname-Ens  & \textbf{0.997} & \textbf{0.985} & \underline{0.961}          & \underline{0.870}           & \underline{0.948}      & \textbf{0.952}    &     {0.988} / 0.009 & 0.781 / 0.01 & 0.486 / 0.009 & 0.427 / 0.01 & 0.689 / 0.008 \\\hline
CANShield-Base & 0.958          & 0.877          & 0.896          & 0.835          & 0.909      & 0.895  & 0.656 / 0.01 & 0.338 / 0.01 & 0.534 / 0.01 & 0.463 / 0.01 & 0.698 / 0.01 \\
CANet\cite{hanselmann2020canet}           & \underline{0.979}          & \underline{0.882}          & \textbf{0.983} & \textbf{0.936} & \textbf{0.974}   & \underline{0.951}  & 0.901 / 0.004 & 0.613 / 0.004 & 0.955 / 0.025 & 0.765 / 0.006 & 0.905 / 0.004 \\
Reconstructive~\cite{weber2018online}     & 0.903          & 0.496           & 0.755            &  0.563          & 0.532      & 0.650  & 0.688 / 0.005 & 0.001 / 0.007 & 0.361 / 0.074 & 0.016 / 0.025 & 0.029 / 0.005\\
Predictive~\cite{taylor2016anomaly}      & 0.874          & 0.489          & 0.722           & 0.561          & 0.530      &  0.635   & 0.644 / 0.006 & 0.003 / 0.007 & 0.33 / 0.026 & 0.015 / 0.006 & 0.02 / 0.004 \\\hline
\end{tabular}%
}}
\end{table*}

\begin{figure*}[!t]
    \centering
  \subfloat[\scriptsize{SynCAN dataset}\label{fig:metric-syncan}]{%
      \includegraphics[width=0.49\linewidth]{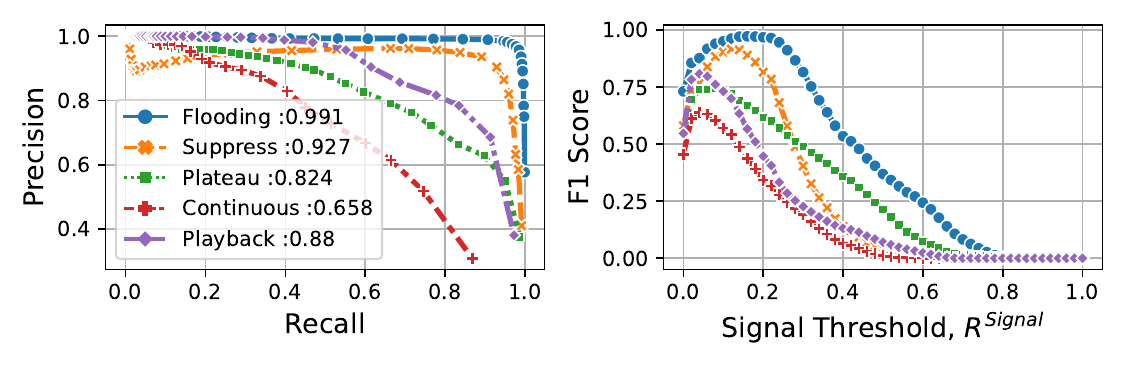}}
    \hfill
  \subfloat[\scriptsize{ROAD dataset}\label{fig:metric-road}]{%
        \includegraphics[width=0.49\linewidth]{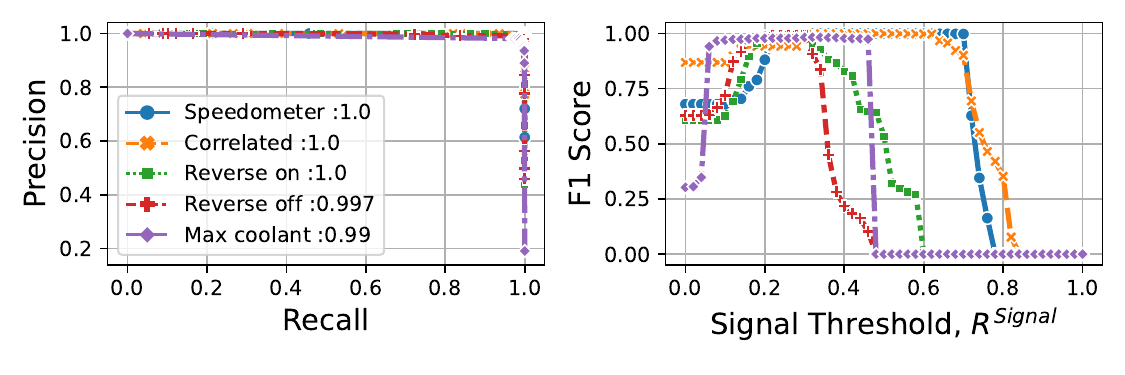}}
    \caption{CANShield-Ens's precision-recall (PR) curve with AUPRC and F1 Scores for different thresholds on both the SynCAN and ROAD datasets.}
    \label{fig:auprc-f1}
\end{figure*}

\vspace{5pt}
\noindent\textbf{ROAD Dataset~}
Fig.~\ref{fig:attack_detect_roc_road} shows the anomaly scores of the attacks on the ROAD dataset. Same as the SynCAN, \sysname-Ens also shows stable performance in the anomaly score. As all the attacks in the ROAD dataset are closely aligned with the \textit{plateau} attack in the SynCAN dataset, both the individual and ensemble models show high performance in detecting the attacks. There are a few cases where the performance degrades a little bit, but \sysname-Ens mitigates such issues and detects all the attacks on the ROAD dataset with an AUROC score of $\sim$ 1.00.

\subsubsection{Precision, Recall, and F1 Score}
In this part, we study the impact of the signal violation thresholds $R^{Signal}$ on \sysname-Ens's precision, recall, and F1 score for different attacks in both the SynCAN and ROAD dataset. The first panel of Fig~\ref{fig:metric-syncan}, which shows the precision-recall curve along with the AUPRC scores on the SynCAN dataset, demonstrates that CANShield-Ens is highly effective against \textit{fabrication} and \textit{suspension} attacks (AUPRC $\ge 0.92$) and moderate performance against advanced \textit{masquerade} attacks (AUPRC$\approx0.65\sim0.88$). Moreover, the values of $R^{Signal}$ within the range of 0.05 to 0.2 provide a decent performance maximizing the F1 scores for different attacks, as shown in the right panel of the figure. Considering CANShield's goal of having a low FPR, we recommend a higher value for $R^{Signal}$, which results in high precision ($>0.9$) for all the attacks.
Similarly, the evaluation results in Fig~\ref{fig:metric-road} on the ROAD dataset show that CANShield-Ens achieves perfect precision, recall, and F1 score (AUPRC$\approx$1.00) with an appropriate threshold ($0.2\sim0.3$).

\vspace{5pt}
\textbf{Comparison with Baseline Models.~}
\label{sec:baseline} 
Whereas we demonstrate the improvements of CANShield-Ens over the individual models, this part includes the performance comparison with the other baseline detectors as well. 
Table~\ref{tab:baseline} illustrates such comparison, which indicates 1.84\% and 11.67\% improvements in the AUROC of \textit{flooding}, and \textit{suppress} attacks, respectively, compared to the closest baseline CANet.
Unlike CANet, CANShield-Ens considers both the sequence of CAN IDs and the time-series signal values to create the data queue and provides effective detection of such practical attacks. Even though CANet performs slightly better against advanced masquerade attacks, 
CANShield-Ens also shows decent performance. \revagain{The right panel of Table~\ref{tab:baseline} shows the TPR and FPR of different CANShield architectures along with the baselines. Similar to the AUROC, CANShield shows promising performance against fabrication and suspension attacks, while CANet performs better against masquerade attacks.} Furthermore, CANShield is considerably lighter than CANet. While CANet consumes 8718 KB of memory~\cite{kukkala2020indra}, CANShield only utilizes 525 KB, making it suitable for edge devices.
\rev{Overall, as Table~\ref{tab:baseline} shows, CANShield-Ens outperforms all of the baselines on average, showing the proposed framework's effectiveness.}


\subsection{Event Detection Latency}
\label{sec:latency}



\begin{figure*}[!t]
    \centering
    \includegraphics[width=0.99\textwidth]{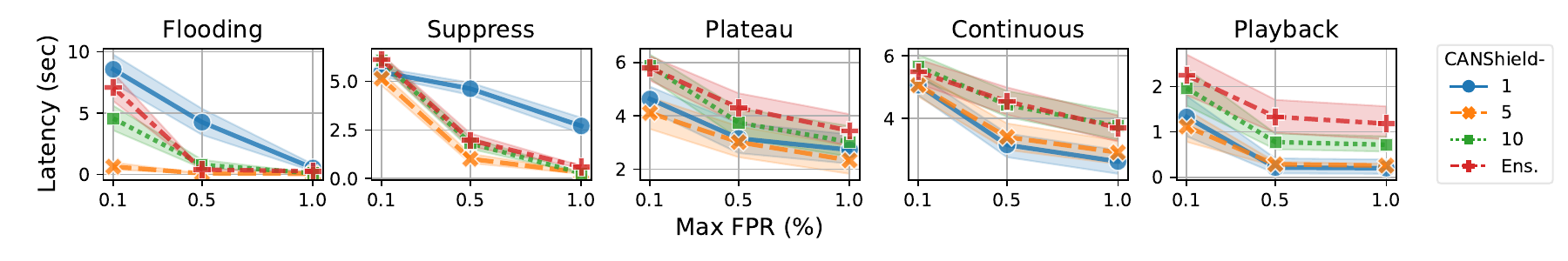}
    \caption{The trade-off between event detection latencies and  maximum FPR thresholds against different attacks in the SynCAN dataset.} 
    \label{fig:event_detect_latency_syncan}
\end{figure*}

Fig.~\ref{fig:event_detect_latency_syncan} illustrates the attack-wise event detection latency for three cases of maximally allowed FPR for the SynCAN dataset. As each attack manipulates the signal at different paces, the time to observe a potential deviation varies. Hence, similar to the previous discussion, certain AEs are more responsive against certain types of attacks. As the first two panels of Fig.~\ref{fig:event_detect_latency_syncan} show that in the case of \textit{fabrication} and \textit{suspension} attacks, CANShield-1 has slightly higher event detection latency, whereas CANShield-Ens reduces the detection latency for the ensemble model. On the other hand, the \textit{masquerade} attacks are the most challenging to detect, and CANShield-Ens reduces the false positives by taking the mean of the final anomaly scores. Therefore, as a trade-off, it increases the latency by a small factor compared to the individual models. However, the latency is still within a small range to cause any devastating impact. It is noted again that in the SynCAN dataset, the attacks were created in post-processing without any physical verification. Hence, some attacks may align with the actual data and lose the malicious property leading to low detection performance and high detection latency.


\begin{figure}[!t]
    \centering
  \subfloat[\scriptsize{Cost of training}\label{fig:tl_cost}]{%
      \includegraphics[width=0.42\columnwidth]{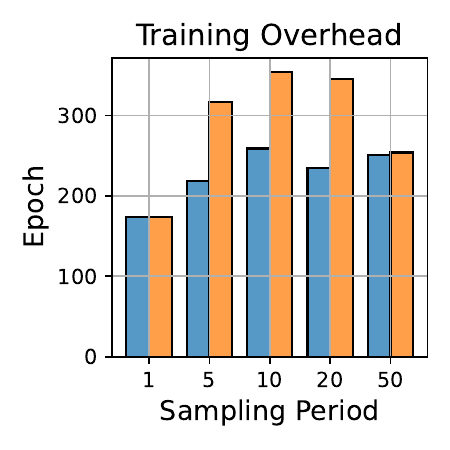}}
    \hfill
  \subfloat[\scriptsize{Validation loss after training}\label{fig:tl_loss}]{%
        \includegraphics[width=0.57\columnwidth]{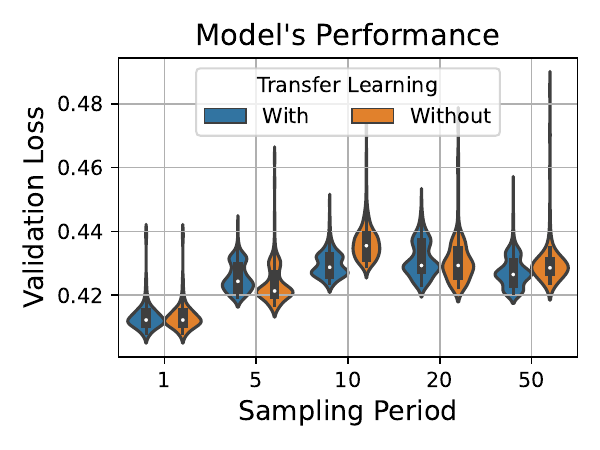}}
\caption{Effectiveness of transfer learning during model training.}
\label{fig:transfer_learning}
\end{figure}


Furthermore, the figures also illustrate the impact of maximum FPR on the event detection latency. Although some individual model suffers from high latency with low FPR (i.e., $0.1\%$), \sysname-Ens provides a lower event detection latency. However, allowing more false positives (max FPR of  $0.5\%-1\%$) into the system further reduces latency. Whereas in case some advanced SynCAN attacks \sysname takes up to a couple of seconds to detect, all the attack events in the ROAD dataset are detected almost instantly (see Fig.~\ref{fig:attack_detect_roc_road}). Therefore, our evaluation shows that \sysname improves detection performance, reduces overall detection latency, and makes the system more robust.




\subsection{Implementation and Processing Latency}
\label{sec:eval-tf}


\vspace{5pt}
\textbf{Transfer Learning.~}
Here, we explain the computational benefit of transferring knowledge from the trained AE models working on lower $T_x$ to AEs with higher $T_x$. Fig.~\ref{fig:tl_cost} shows that without any knowledge transfer, the number of training epochs to reach the early stopping criteria, which is a steady validation loss, increases by up to 100\% of the initial training
for different AEs. However, if the AE model's parameters are initialized as the pre-trained AE with the immediate lower $T_x$, the number of training epochs gets reduced by approximately 30\% in most cases. Besides, as Fig.~\ref{fig:tl_loss} shows, such initialization does not impact the performance of the final models as the validation loss of the final AE models remains almost the same regardless of the weight initialization. Therefore, CANShield-Ens reduces the training cost of consecutive AEs significantly by transferring the weights to the next AE without any performance trade-off.

\vspace{5pt}
\textbf{Hardware Processing Latency.~} We trained and evaluated \sysname on a laptop with a 2.3 GHz 8-Core Intel i9 processor with 32 GB of RAM and AMD Radeon Pro 5500M 8 GB of graphics and also deployed on a Raspberry Pi with 1.5GHz 64-bit quad-core CPU and 4GB of RAM to benchmark \sysname prediction speed. To reduce the inference time and the size of AE models, we convert the TensorFlow model into TensorFlow Lite~\cite{tf-lite} models, which quantizes the weights. Results show each \sysname process takes around $1$ms on the laptop, which satisfies our design objective ($<2$ms), and \add{$10$ms on the Raspberry Pi, which is low for an attack to cause catastrophe to the targeted vehicle.} \revagain{Our extensive testing and validation demonstrate that the quantized AE-based CANShield provides no degradation in performance and yields the same detection results as the original ones.}

\rev{\subsection{Limitations and Discussions}

Here, we discuss two key challenges of \sysname, which are common for any DL-based signal-level CAN IDS. 
\begin{itemize}
    \item The first challenge is to get the DBC files from the OEM or have an efficient reverse engineering tool to create the signal-level representation of the CAN dataset. Hence, we assume that the defender is OEM who has direct access to the DBC file or a third party with an efficient reverse engineering tool.
    \item The collection of sufficient training data and generalizing the training of the AE models is another challenge. 
    To overcome these issues, CANShield is assumed to be trained on a very dynamic high-fidelity dataset, including a diverse range of driving patterns and various driving scenarios, to ensure that it can detect anomalies regardless of the driving context and driver's behavior.
\end{itemize}}




\section{RELATED WORK}
\label{sec:related_work} 
There has been a good amount of work on CAN IDS, which can be divided into the following general categories.

\noindent\textbf{Physical Characteristics-based IDS.~} One line of research in CAN IDS utilized the physical layer attributes of the CAN bus communications to fingerprint the ECUs and verify the source of each message. Since the physical signals generated from the ECUs solely depend on the ECUs' hardware characteristics, it is assumed to be unique; hence, a malicious ECU cannot controllably modify it. Therefore, such defense has been considered effective in detecting injection attacks. Out of different attributes, clock skews~\cite{cho2016fingerprinting}, voltage profile~\cite{cho2017viden,choi2018voltageids}, electrical CAN signal characteristics~\cite{choi2018identifying, kneib2018scission}, etc. are widely used in fingerprinting and building physical characteristics-based IDS. 
However, the assumption of the uniqueness of such physical properties is proven invalid by a recent study~\cite{bhatia2021evading}, which proposed a voltage corruption tactic that can modify the physical attributes of the victim ECU and impersonate the targeted ECU. Therefore, such IDSs cannot provide a comprehensive security guarantee against a wide range of cyberattacks.

\noindent\textbf{CAN ID-based IDS.~} A vast portion of the attacks, especially \textit{fabrication} and \textit{suspension} attacks, consider exploiting the sequences of CAN IDs to disrupt regular services. Therefore, some IDSs extract features from the series of CAN IDs to learn the usual pattern and detect abnormalities.
Given the labeled datasets, some works utilized different types of supervised learning models, based on CNN~\cite{song2020vehicle, desta2022rec}, long short-term memory (LSTM)~\cite{jedh2021detection}, support vector machine, k-nearest neighbors, decision tree, random forest, and XGBoost~\cite{refat2021detecting, han2021event, aliyu2021blockchain} etc., to build the IDSs. 
Different unsupervised ML algorithms are also studied in CAN ID-based IDS research. Various features, such as message timing information per CAN ID and window-wise ID-counting, are used as the underlying features for the IDSs~\cite{blevins2021time}.

A few works predicted the next CAN ID with individual LSTM or gated recurrent unit (GRU) models and used log loss and a predefined threshold to detect malicious injections~\cite{rajapaksha2022keep}. 
Similarly, one-class support vector machine (OCSVM)~\cite{al2019intelligent}, isolation forest~\cite{sharmin2021intrusion} are also studied. Along with unsupervised methods, self-supervised method-based IDS are also studied~\cite{song2021self}. A few works converted the sequences of CAN IDs into 2D images and trained generative adversarial networks (GANs) in an unsupervised fashion~\cite{seo2018gids, zhao2022can}. 
Recently, motivated by natural language processing, some researchers considered the sequence of CAN IDs as a sentence and utilized world embedding and language models to build the CAN IDS~\cite{shi2021intrusion, baldini2021intrusion}. 
The fundamental drawback of the CAN ID-based IDSs is that they are only effective against injection attacks that explicitly change the sequence of IDs. However, advanced masquerade attacks can manipulate the payload without disrupting the ID sequences/frequencies and easily evade such IDSs~\cite{miller2019lessons}.  

\noindent\textbf{Payload-based Detection.~}
The advanced attacks can not only change the CAN IDs but also modify the payloads of the messages. The attacker can replay prerecorded values or change the actual values. Hence, there has been a good amount of work learning the pattern in the payload sequences and using it to detect potential cyberattacks.
Extracting usable features from the binary payloads is a challenging task. The mode and value information is commonly used to extract features and implement DNN-based IDS~\cite{kang2016intrusion}. 
A few works proposed a continuous field classification (CFC) algorithm to identify the payload value alignments and used a deep learning-based approach to identify the anomalous fields~\cite{fenzl2020continuous}. Moreover, different k-nearest neighbor classifiers are also used to identify different attacks~\cite{martinelli2017car}.  
Considering the sequence of CAN messages as time series data, a few works implemented unsupervised ML models based on  LSTM~\cite{wang2020vulnerability, khan2019vehicle} and OCSVM to build the payload-based CAN IDS~\cite{chockalingam2016detecting}. 

\noindent\textbf{Signal-level Detection.~} Compared to the IDSs mentioned above, IDSs working at the time-series signal-level can extract the most useful information and build an efficient and context-aware decision model. Moriano et al.~\cite{moriano2022detecting} hypothesized that masquerade attacks alter the correlations among the signals and the clustering behaviors and proposed a technique to detect such attacks by comparing the clustering similarity of test data with and without attack traces. Recent works proposed DNN-based signal-level CAN IDS, where the extracted sensor values are used as separate features for IDS~\cite{zhang2019intrusion}. 
Other research efforts also proposed RNN/LSTM-based models with an embedding layer working on CAN payload values in~\cite{taylor2016anomaly, tanksale2020anomaly,  ashraf2020novel}. A few similar approaches in CAN IDS research used GRU, LSTM, and temporal CNN-based AEs for each CAN ID~\cite{tanksale2020anomaly,  ashraf2020novel, kukkala2020indra, longari2020cannolo, kukkala2021latte, thiruloga2022tenet}. All of these IDSs~\cite{tanksale2020anomaly,  ashraf2020novel, kukkala2020indra, longari2020cannolo, kukkala2021latte, thiruloga2022tenet} processed ID-wise data independently and utilized individual models for each ID, which ignored the signal-wise correlations and fail to detect attack collectively. 

CANet~\cite{hanselmann2020canet} is one of the closest works to our proposed method. It employed one LSTM model for the signals with each CAN ID, 
and used AE-based reconstruction to predict the anomaly score. However, in practice, LSTM networks are costly to train, and one LSTM for each IDS will make it impractical for a vehicle with many CAN IDs. 
Moreover, due to the complicated architecture, CANet shows low detection performance on \textit{suppress} attacks, a form of the well-known bus-off attack that can be easily launched due to the CAN protocol's limitations. In~\cite{novikova2020autoencoder}, the authors proposed to manually group the highly correlated signals into smaller subgroups and use AE for each subgroup. However, such manual clustering is not feasible for real vehicles with lots of signals.


\section{CONCLUSION}
\label{sec:conclusion}
As modern vehicles become more connected to external networks, the attack surface of the CAN bus system grows drastically. 
To secure the CAN bus from advanced intrusion attacks, we propose a signal-level intrusion detection framework, \sysname. 
With the capability of handling a high-dimensional CAN data stream, \sysname trains multiple CNN-based AE models to work on different views of the data stream across different temporal scales, performs a three-step structural analysis of the reconstruction losses, and finally ensembles them to obtain the final anomaly score. Evaluation results on both the SynCAN and ROAD datasets show CANShield's robustness and responsiveness against different advanced attacks. The proposed three-step analysis of the reconstruction loss improves the overall AUROC by 6.40\% than the conventional mean average method. The aggregation of data with different temporal scales reduces variance in inference and increases the AUROC by at least 2.19\% compared to any single AE-based framework. 
\revagain{Moreover, CANShield outperforms all the baselines against practical fabrication and suspension attacks while still performing well against advanced masquerade attacks. By combining the strengths of CAN ID-based IDS and signal-level IDS, CANShield offers a scalable and efficient solution and advances the state-of-the-art.}
\section*{ACKNOWLEDGMENT}
This work was supported in part by the US National Science Foundation (NSF) under NSF grants CNS-1837519 and CNS-2235232, the Virginia Commonwealth Cyber Initiative (CCI), and the Laboratory Directed Research and Development Program of Oak Ridge National Laboratory (ORNL), managed by UT-Battelle, LLC, for the U.S. Department of Energy. We are also thankful to Robert A. Bridges from ORNL for his insightful comments on the manuscript.
\bibliographystyle{unsrt}
\bibliography{reference}
\begin{IEEEbiography}[{\includegraphics[width=1in,height=1.25in,clip,keepaspectratio]{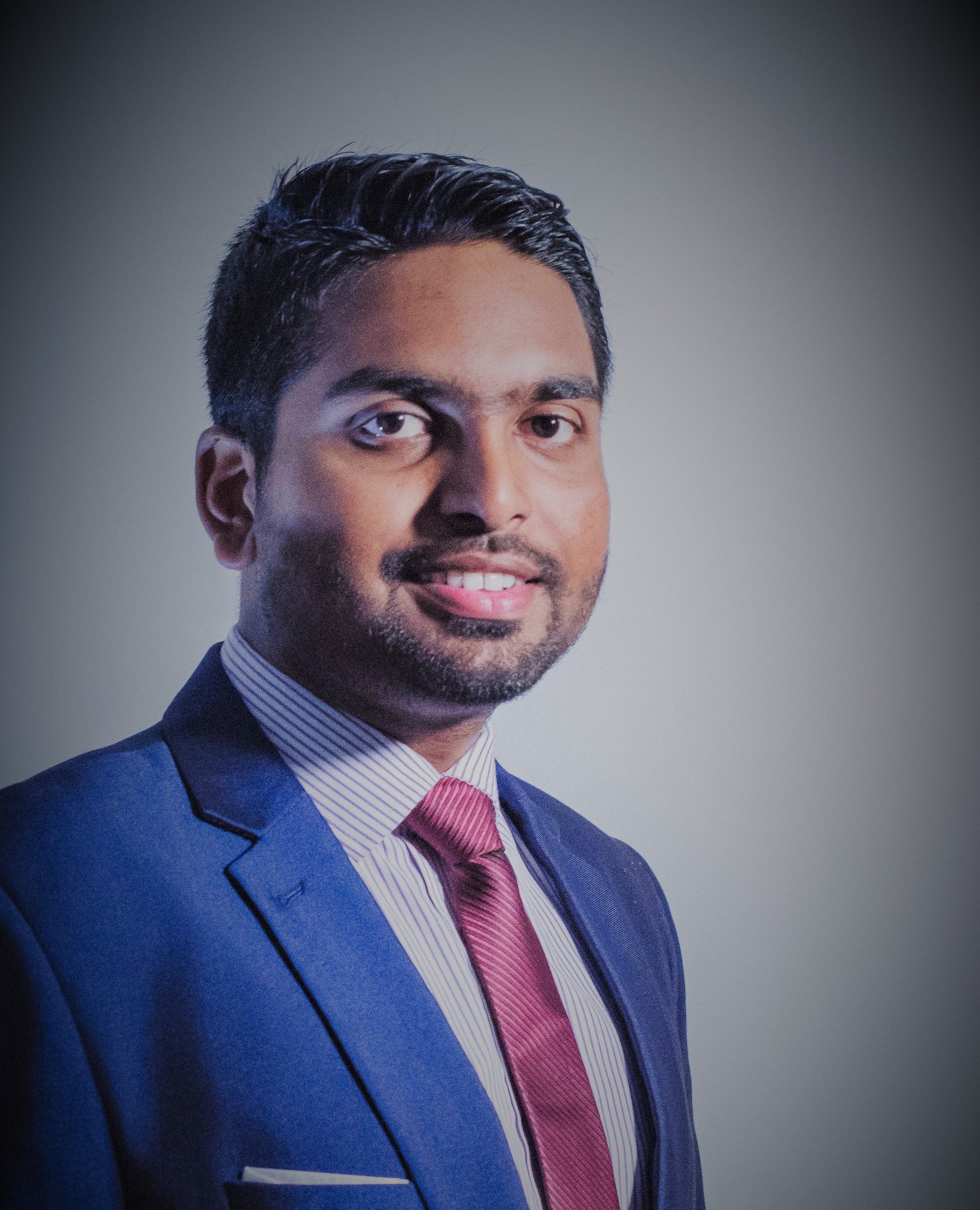}}]
{Md Hasan Shahriar}~(Student Member, IEEE) is currently pursuing a Ph.D. degree in Computer Science at Virginia Tech under the supervision of Prof. Wenjing Lou. He received a B.Sc. degree in Electrical and Electronic Engineering from Bangladesh University of Engineering and Technology in 2016 and an M.S. degree in Computer Engineering from Florida International University in 2020. His research interests include automotive cybersecurity, cyber-physical systems, and machine learning.
\end{IEEEbiography}

\begin{IEEEbiography}[{\includegraphics[width=1in,height=1.25in,clip,keepaspectratio]{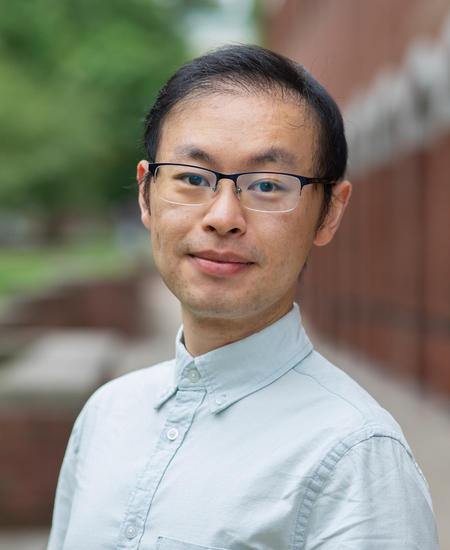}}]{Yang Xiao}~(Member, IEEE) is an Assistant Professor in the Department of Computer Science, University of Kentucky. He received his Ph.D. degree from the Bradley Department of Electrical and Computer Engineering at Virginia Tech, supervised by Prof. Wenjing Lou. He received
his B.S. degree from the School of Electrical and Information Engineering at Shanghai Jiao Tong University and his M.S. degree from the Electrical Engineering and Computer Science Department at the University of Michigan, Ann Arbor. His research interests lie in network security, blockchain and distributed system security, and wireless network security.
\end{IEEEbiography}

\begin{IEEEbiography}[{\includegraphics[width=1in,height=1.25in,clip,keepaspectratio]{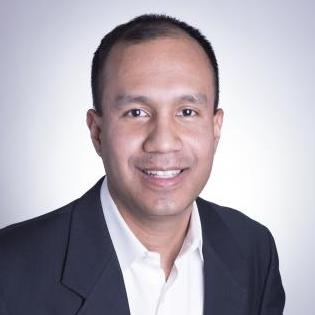}}]{Pablo Moriano}~(Senior Member, IEEE) (SM'21) is a research scientist in the Computer Science and Mathematics Division at Oak Ridge National Laboratory. He received Ph.D. and M.S. degrees in Informatics from Indiana University Bloomington. Previously, he received M.S. and B.S. degrees in Electrical Engineering from Pontificia Universidad Javeriana in Colombia. Dr. Moriano's research lies at the intersection of data science, network science, and cybersecurity. In particular, he uses data-driven and computational methods to discover and understand critical security issues in large-scale networked systems. Applications of his research range across multiple disciplines, including, the detection of exceptional events in social media, Internet route hijacking, and insider threat behavior in version control systems. He is a senior member of IEEE, and a member of ACM and SIAM.

\end{IEEEbiography}

\vspace{-12pt}
\begin{IEEEbiography}[{\includegraphics[width=1in,height=1.25in,clip,keepaspectratio]{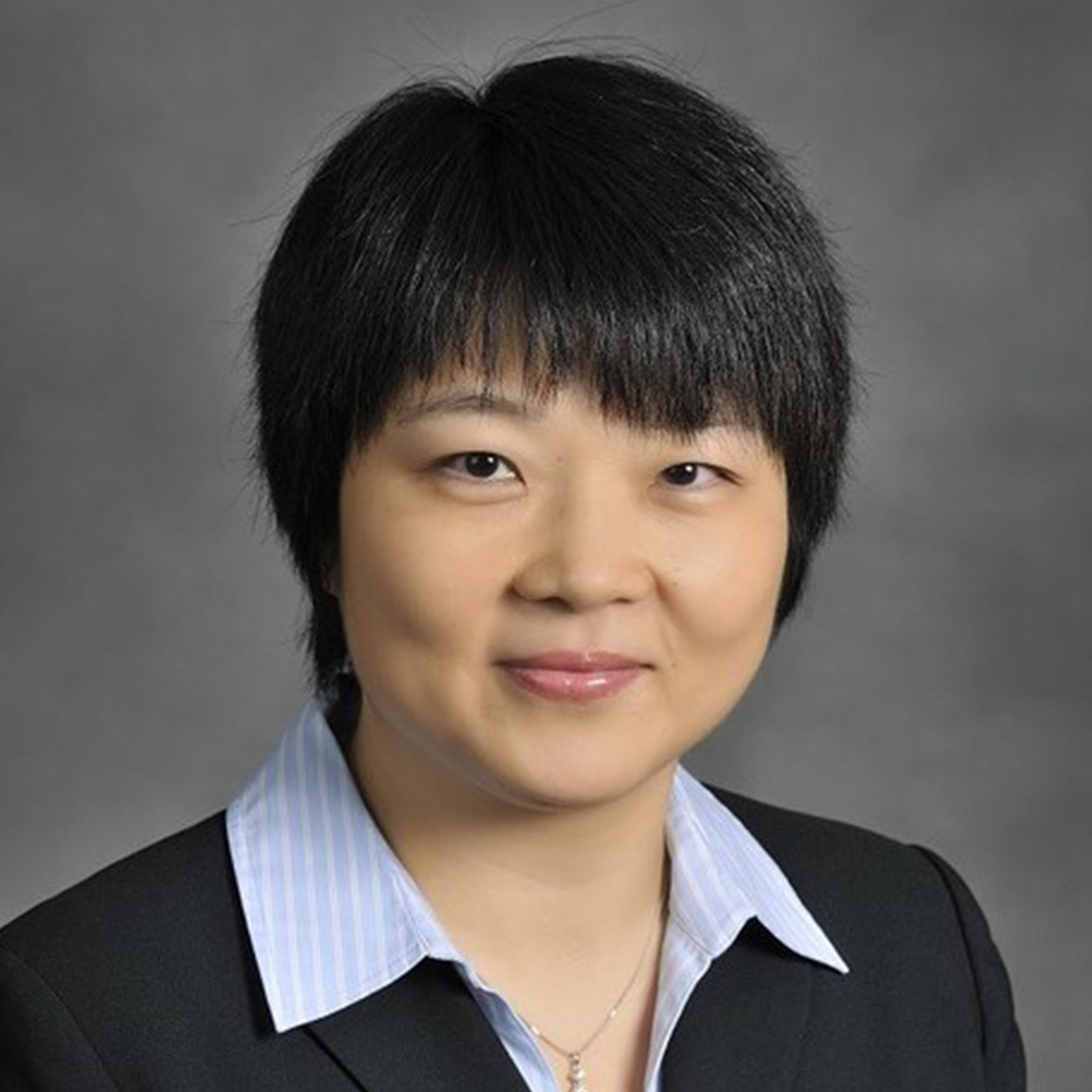}}]{Wenjing Lou}~(Fellow, IEEE) (F’15) received the Ph.D. degree in electrical and computer engineering from the University of Florida in 2003. She is currently the W. C. English Endowed Professor of Computer Science at Virginia Tech and a Fellow of the IEEE. Her research interests cover many topics in the cybersecurity field, with her current research interest focusing on wireless network security, trustworthy AI, blockchain, and security and privacy problems in the Internet of Things (IoT) systems. Prof. Lou is a highly cited researcher by the Web of Science Group. She received the Virginia Tech Alumni Award for Research Excellence in 2018. She received the INFOCOM Test-of-Time paper award in 2020. She was the TPC chair for IEEE INFOCOM 2019 and ACM WiSec 2020. She was the Steering Committee Chair for IEEE CNS conference from 2013 to 2020. She is currently a steering committee member of IEEE INFOCOM and IEEE Transactions on Mobile Computing. She served as a program director at the US National Science Foundation (NSF) from 2014 to 2017.
\end{IEEEbiography}

\begin{IEEEbiography}[{\includegraphics[width=1in,height=1.25in,clip,keepaspectratio]{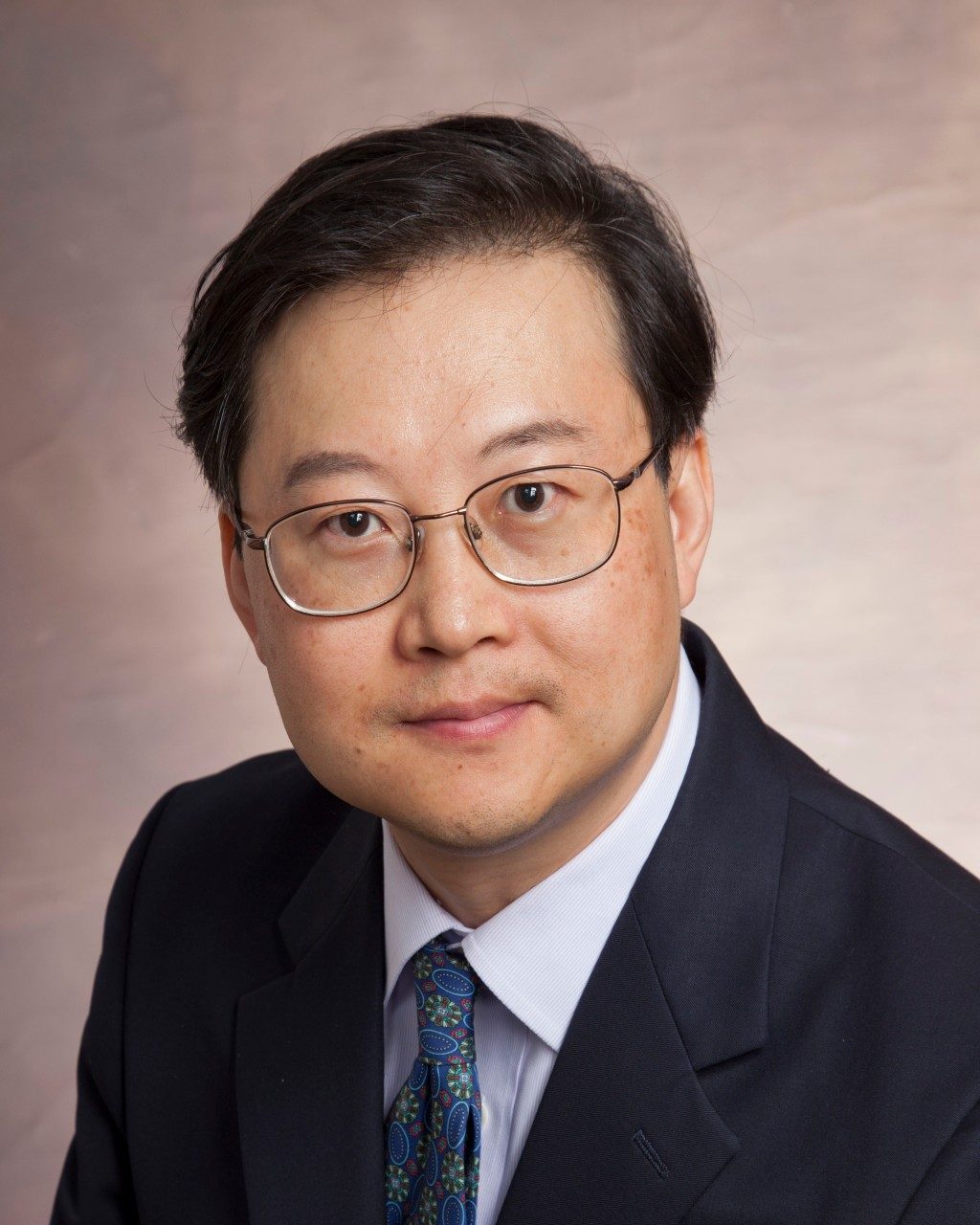}}]{Y. Thomas Hou}~(Fellow, IEEE) (F’14) received his Ph.D. from NYU Tandon School of Engineering in 1998. He is currently Bradley Distinguished Professor of Electrical and Computer Engineering at Virginia Tech, Blacksburg, VA, USA, which he joined in 2002. He was a Member of Research Staff at Fujitsu Laboratories of America in Sunnyvale, CA from 1997 to 2002. 
His current research focuses on developing innovative real-time solutions to complex science and engineering problems arising from wireless and mobile networks. 
He is also interested in wireless security. He has published over 350 papers in IEEE/ACM journals and conferences. His papers were recognized by 10 best paper awards from IEEE and ACM, including an IEEE INFOCOM Test of Time Paper Award in 2023. He holds six U.S. patents. He authored/co-authored two graduate textbooks: {\em Applied Optimization Methods for Wireless Networks \/}(Cambridge University Press, 2014) and {\em Cognitive Radio Communications and Networks: Principles and Practices\/} (Academic Press/Elsevier, 2009). Prof.~Hou was named an IEEE Fellow for contributions to modeling and optimization of wireless networks. He was/is on the editorial boards of a number of IEEE and ACM transactions and journals. He was Steering Committee Chair of IEEE INFOCOM conference and was a member of the IEEE Communications Society Board of Governors.  He was also a Distinguished Lecturer of the IEEE Communications Society.  
\end{IEEEbiography}
\vfill
\pagebreak

\vfill
\end{document}